\documentclass[useAMS,usenatbib]{mn2e}
\title[An RV survey for binary $\lbrack$WR$\rbrack$ CSPNe]{A
radial velocity survey for post-common-envelope Wolf-Rayet central
stars of planetary nebulae: First results and discovery of the close
binary nucleus of NGC~5189\thanks{Based on observations made with the
South African Astronomical Observatory (SAAO) 1.9-m telescope and the
Southern African Large Telescope (SALT) under programme 2013-2-RSA-005
(PI: Miszalski).}}
\author[Manick, Miszalski \& McBride]
{Rajeev Manick,$^{1,2,3}$\thanks{E-mail: rajeev@ster.kuleuven.be}
  Brent Miszalski$^{1,4}$
  and Vanessa McBride$^{1,2}$\\
$^{1}$South African Astronomical Observatory, PO Box 9, Observatory,
7935, South Africa\\
  $^{2}$Astrophysics, Cosmology and Gravity Centre, Department of
Astronomy, University of Cape Town, Private Bag X3,\\
       Rondebosch 7701, South Africa\\
$^{3}$Instituut voor Sterrenkunde, KU Leuven, Celestijnenlaan
  200D bus 2401, B-3001 Leuven, Belgium\\
$^{4}$Southern African Large Telescope Foundation, PO Box 9,
Observatory, 7935, South Africa\\}
\date{Released 2014 Xxxxx XX}

\pagerange{\pageref{firstpage}--\pageref{lastpage}} \pubyear{2002}

\def\LaTeX{L\kern-.36em\raise.3ex\hbox{a}\kern-.15em
    T\kern-.1667em\lower.7ex\hbox{E}\kern-.125emX}
\usepackage{graphicx}
\usepackage{color}
\usepackage{calc}

\newcommand{\aj}{AJ}
\newcommand{\apj}{ApJ}
\newcommand{\apjl}{ApJ}
\newcommand{\apss}{Ap\&SS}

\newcommand{\apjs}{ApJS}

\newcommand{\aaps}{A\&A S}

\newcommand{\rmxaa}{RMxAA}
\newcommand{\nar}{NAR}
\newcommand{\aap}{A\&A}

\newcommand{\araa}{ARA\&A}
\newcommand{\mnras}{MNRAS}
\newcommand{\pasp}{PASP}

\begin{document}
\date{Accepted Received ; in original form }

\pagerange{\pageref{firstpage}--\pageref{lastpage}} \pubyear{2014}
\label{firstpage}

\maketitle

\begin{abstract}
   The formation of Wolf-Rayet central stars of planetary nebulae ([WR] CSPNe) whose spectroscopic appearance mimics massive WR stars remains poorly understood. Least understood is the nature and frequency of binary companions to [WR] CSPNe that may explain their H-deficiency. We have conducted a systematic radial velocity (RV) study of 6 [WR] CSPNe to search for post-common-envelope (post-CE) [WR] binaries. We used a cross-correlation method to construct the RV time-series as successfully done for massive close binary WR stars. No significant RV variability was detected for the late-[WC] type nuclei of Hen~2-113, Hen~3-1333, PMR~2 and Hen~2-99. Significant, large-amplitude variability was found in the [WC4] nucleus of NGC~5315. In the [WO1] nucleus of NGC~5189 we discovered significant periodic variability that reveals a close binary with $P_\mathrm{orb}=4.04\pm0.1$ d. We measured a semi-amplitude of $62.3\pm1.3$ km s$^{-1}$ that gives a companion mass of $m_2\ge0.5$ $M_\odot$ or $m_2=0.84$ $M_\odot$ (assuming $i=45^\circ$). The most plausible companion type is a massive WD as found in Fleming~1. The spectacular nebular morphology of NGC~5189 fits the pattern of recently discovered post-CE PNe extremely well with its dominant low-ionisation structures (e.g. as in NGC~6326) and collimated outflows (e.g. as in Fleming~1). The anomalously long 4.04 d orbital period is either a once-off (e.g. NGC~2346) or it may indicate there is a sizeable population of [WR] binaries with massive WD companions in relatively wide orbits, perhaps influenced by interactions with the strong [WR] wind.
\end{abstract}


\section{Introduction}

Planetary nebulae (PNe) are circumstellar gas envelopes ejected at the
end of the Asymptotic Giant Branch (AGB) phase of low-intermediate mass stars of 
$\sim$$1M_{\odot}$ to $8M_{\odot}$. A majority of central stars of planetary
nebulae (CSPNe) show hydrogen rich atmospheres (e.g. M\'endez et al.
1988; M\'endez 1991; Napiwotzki \& Schoenberner 1995), with their
spectra containing only weak absorption lines, mainly of hydrogen (H)
and helium (He). Perhaps the least understood are the less numerous
emission-line CSPNe with H-deficient atmospheres whose spectral
appearance mimics the massive Wolf-Rayet (WR) stars. These CSPNe
exhibit dense and strong stellar winds with mass-loss rates up to
$\sim$10$^{-7}$ M$_\odot$ yr$^{-1}$ (e.g. Koesterke 2001; Crowther
2008). Their classification is similar to massive WR stars where the
dominant emission lines indicate either a [WC] (He, C and O; Crowther
et al. 1998) or [WN] (He and N; Smith et al. 1996) subtype, where the
brackets around the spectral type distinguish them from their
massive counterparts. In contrast to the roughly equal division
between WN and WC subtypes in massive WR stars (van der Hucht 2001),
the known examples of [WR] stars are heavily skewed towards early and
late [WC] types with few intermediate [WC] types (Crowther 2008; see
however G\'orny 2014), whereas only two bona-fide [WN] CSPNe were established
recently in IC~4663 (Miszalski et al. 2012) and A~48 (Todt et al.
2013; Frew et al. 2014). Comprehending this disparity may shed much needed light on how [WR] stars form.

The most peculiar aspect of the newly discovered [WN] CSPNe is that
their atmospheres are extremely helium rich (85--95 per cent by mass,
Miszalski et al. 2012; Todt et al. 2013). Therefore, they do not fit
into the C-rich evolutionary sequence that the vast majority of
H-deficient post-AGB stars follow (Werner \& Herwig 2006). The
C-rich H-deficient post-AGB evolutionary sequence of
[WC] stars evolving into PG1159 stars, the latter characterised by significantly weaker winds and higher surface gravities, is well supported by similar atmospheric abundance
patterns between the two classes (Crowther 2008; Werner et al. 2008).
Werner (2012) suggested that [WN] CSPNe `might' be the progenitors of
O(He) stars, namely a rare group of four high surface gravity He-rich
post-AGB stars, two of which are surrounded by PNe (Rauch et al. 1994,
1996, 1998, 2008; Reindl et al. 2013, 2014). However, the existence of
[WN] CSPNe was still unproven at this time and PB~8 was not He-rich
enough to be strictly comparable to the O(He) stars. Only with the
discovery of the extremely He-rich atmosphere of the [WN3] CSPN of
IC~4663 (95 per cent He by mass) could the O(He) stars, which are
similarly He-rich with He mass fractions $\ga$90 per cent (Reindl et
al. 2013, 2014), be securely linked to [WN] CSPNe in a separate
He-rich H-deficient post-AGB evolutionary sequence [WN]$\to$O(He) by Miszalski et al. (2012). In this new sequence the O(He) stars are the He-rich analogues of the C-rich PG1159 stars.

A key question prompted by the discovery of the He-rich sequence is
whether it operates in a similar fashion to the C-rich sequence.
Is the appearance of the [WR] phenomenon produced by the same recipe?
Or is the phenomenon produced via separate routes with different
recipes? There are compelling arguments to include binary
interactions in the formation of some [WC] CSPNe and related objects
(De Marco \& Soker 2002; De Marco 2008) for which the standard
formation scenarios of a late thermal pulse (LTP) (e.g. Schoenberner
1979; Bl{\"o}cker 2001; Herwig 2001), a very late thermal pulse (VLTP)
(e.g. Lawlor \& MacDonald 2002), or an AGB final thermal pulse (AFTP)
(e.g Herwig 2001) may not apply.\footnote{These scenarios cannot reproduce the
moderate level of $\sim$55 per cent He by mass in the possible
intermediate [WN/WC] CSPN of PB~8 (Todt et al. 2010). Miller Bertolami
et al. (2011) proposed a diffusion induced nova scenario to explain
the abundance pattern of PB~8, however these models require low
metallicities and core masses to reach the extremely He-rich
compositions of [WN] CSPNe (see the discussion in Todt et al. 2013). 
While it seems that low metallicities may facilitate [WR] formation (Zijlstra et al. 2006;
Kniazev et al. 2008), the $\alpha$-element nebular abundances of both
IC4663 and A~48 are notably Solar, suggesting the Miller Bertolami
et al. (2011) scenario may not be applicable.} 
Miszalski et al. (2012) suggested a merger may be responsible
for the high He enrichment in IC~4663 and Reindl et al. (2013) found a 
reasonable agreement between the atmospheric
composition of IC~4663 and that predicted from a slow merger of two He-WDs (Zhang \& Jeffery 2012).
A merger scenario would be consistent with the extreme rarity of both
[WN] and O(He) stars compared to their C-rich counterparts,
however it may not be possible to prove observationally.
Some other scenarios were discussed by Reindl et al. (2014), however we
emphasise that many of the scenarios require considerably more
theoretical and, in particular, observational investigation.

Unfortunately, there is currently insufficient evidence to gauge
whether binary interactions in H-deficient post-AGB stars are
widespread. Are the known binaries merely an oddity where the [WR]
component formed independently of binary evolution? Or was a binary
interaction fundamental to forming the [WR] component?

Several different possibilities for binary interactions were proposed
by De Marco \& Soker (2002) and De Marco (2008), however the paucity of
observed binary systems makes it difficult to discern to what degree
these or some other scenarios may play a role. 
In this work we aim to address this by searching for companions in a sample of [WR] central stars. 
It is more practical with modest amounts of telescope time to search for close binaries 
where the orbital periods are a few days or less,
signifying that a common-envelope (CE) interaction has produced the observed short orbital periods 
(Iben \& Livio 1993; Ivanova et al. 2013). Only two previous post-CE examples have been
discovered, namely SDSS J212531.92$-$010745.9 that consists of a
PG1159 star with an M dwarf companion (Nagel et al. 2006; Schuh et al.
2009) and the [WC7] CSPN of PN~G222.8$-$04.2 with an undetermined
companion type (Hajduk et al. 2010). Their literature orbital periods
are 0.29 d and 0.63 or 1.26 d, respectively. 

The majority of close binary CSPNe have been found via photometry monitoring including PN~G222.8$-$04.2.\footnote{See Table~\ref{tab:closebinaries} in Appendix for an updated compilation.} There are however several limitations with this approach. A notable exception is the double degenerate 
nucleus of Fleming~1 (Boffin et al. 2012), discovered from radial velocity (RV) monitoring. The companion in Fleming~1 is a more massive WD whose presence is not detectable in the optical lightcurve. At least 37 unique early-type [WC], PG~1159 and O(He) 
CSPNe have been subjected to photometric monitoring campaigns to search for non-radial pulsations (Ciardullo \& Bond 1996; Gonzalez-Perez et al. 2006), however no close binaries have been discovered as part of these campaigns. If $\sim$15-20 per cent of 
PNe are expected to host close binaries (Bond 2000; Miszalski et al. 2009a), then we would have expected $\sim$5--7 close binaries amongst the sample. It is particularly suprising that no binaries were found, especially since they are amongst 
the hottest CSPNe known and as such they would have produced a very large irradiation effect if a close main sequence companion was present (e.g. Corradi et al. 2011; Miszalski et al. 2011). If these CSPNe were to have companions, they are most likely 
evolved, massive WD companions as in Fleming~1 (Boffin et al. 2012). Another possibility is that there are main sequence companions present with orbital periods of several days or months, for which an irradiation effect would not be detected (e.g. De Marco et al. 2008). 
An RV monitoring study is therefore the least biased method to search for elusive [WR] close binary CSPNe, however such a study has not been previously attempted. Previously researchers have focused primarily on Of-type CSPNe, whose photometry 
(e.g. Handler 2003) and line profiles (e.g. M\'endez 1990; De Marco et al. 2004; De Marco 2009) show significant variability. 
Perhaps one reason these studies have not been extended to [WR] CSPNe is that they were presumed to be similarly affected by wind variability (if not more so for those with faster wind speeds), however several studies of 
massive WR stars have routinely found close binaries (Foellmi et al. 2003; Sana et al. 2013; Schnurr 2008; Schnurr et al. 2009) and there is 
no reason to suspect the same methods cannot be applied to [WR] CSPNe.

This paper is structured as follows: Section~\ref{sec:sec2} describes our sample selection and the observations. Section~\ref{sec:sec3} gives an overview of the cross-correlation method used to find RV shifts in the spectra. 
In Section~\ref{sec:sec4} we present our results 
for each object and we perform a $\chi^2$ analysis to quantify variability. Section \ref{sec:sec5} derives the orbital period and mass function of the most variable object, NGC~5189, before discussing the nature of the companion and the nebular morphology. We conclude in Section \ref{sec:concl}.

\section{Observations} \label{sec:sec2}

\subsection{Sample selection}
Table \ref{tab:sample} gives the basic properties of the objects considered in this pilot study. All were monitored using the cassegrain spectrograph (SpCCD) 
on the South African Astronomical Observatory (SAAO) 1.9-m telescope except for NGC~5189 which was observed with the Robert Stobie Spectrograph 
(RSS; Burgh et al. 2003; Kobulnicky et al. 2003) on the queue-scheduled Southern African Large Telescope (SALT; Buckley, Swart \& Meiring 2006; O'Donoghue et al. 2006). 
In general we required the spectra to have a resolution of $\la$2\AA\ (full-width at half maximum, FWHM) and S/N $\ga$ 40 in the continuum to be able to ensure cross-correlation techniques 
could achieve a velocity accuracy of $\sim$10 km s$^{-1}$ or better. Given the relatively poor throughput of SpCCD, we were heavily restricted in our choice 
of [WR] CSPNe to monitor, namely those brighter than $V\la14$ mag. Therefore we selected several suitably bright targets visible at the time of observations 
from Acker \& Neiner (2003) and one from Morgan et al. (2001). The magnitude restrictions also imposed a bias towards intrinsically brighter late [WC] CSPNe. 
The sample observed by the SAAO 1.9-m is not particularly remarkable, apart from the peculiar morphologies of Hen~2-113 (Lagadec et al. 2006) and Hen~3-1333 
(Chesneau et al. 2006). Both Hen~2-113 and Hen~3-1333 exhibit dual-dust chemistry which may have originated from binary evolution. A binary system may be 
necessary to explain the resolved dust disk in Hen~3-1333 (De Marco et al. 2002; Chesneau et al. 2006) as they are a prevailing feature of post-AGB binaries 
(Van Winckel et al. 2009). The inclusion of NGC~5189 was specially motivated for SALT observations as described in the following.

\begin{table*}
\centering
\caption{Basic properties of the sample of [WR] CSPNe observed.}
\label{tab:sample}
\begin{tabular}{llllllrl}
\hline \hline
PNG  &  Name  & RA & Dec. & $V$ & Type & Epochs & Ref. \\ \hline
291.3$+$08.4  & PMR~2  &  11 34 38.6 & $-$52 43 33 & 13.3 & [WC9/10]& 7 & c \\
307.2$-$03.4  & NGC~5189 & 13 33 32.8 & $-$65 58 27 & 14.5 & [WO1] & 14 & b,d\\
309.0$-$04.2  & Hen~2-99  &  13 52 30.7 & $-$66 23 26 & 13.3 & [WC9]& 7 & a,b \\
309.1$-$04.3  & NGC~5315  &  13 53 57.1 & $-$66 30 51 & 14.4 & [WC4] & 4 & a,b \\
321.0$+$03.9  & Hen~2-113  &  14 59 53.5 & $-$54 18 07 & 11.9 & [WC10] & 12  & a,b \\
332.9$-$09.9  & Hen~3-1333  &  17 09 00.9 & $-$56 54 48 & 10.9 & [WC10] & 7 & a,b \\
\hline
\end{tabular}
\begin{flushleft}
References: (a): Acker and Neiner (2003).
(b): Crowther et al. (1998).
(c): Morgan et. al. (2001).
(d): Ciardullo et al. (1999).
\end{flushleft}

\end{table*}

NGC~5189 (PN G307.2$-$03.4) is a bright, remarkably peculiar southern PN that has long perplexed astronomers. 
It was discovered in 1826 by J. Dunlop from Australia (Cozens, Walsh \& Orchiston 2010), but its unusual appearance 
meant that it was not until much later that it was recognised to be a PN with a `quasi-planetary' classification assigned 
by Henize (1967). The filamentary nebula has been described as `a graceful affair of recurvant gaseous draperies' (Evans 1968) 
and its notorious complexity is the result of multiple point-symmetric low-ionisation knots or ansae (Phillips \& Reay 1983; Hua, 
Dopita \& Martinis 1998; Gon{\c c}alves et al. 2001; Sabin et al. 2012). Phillips \& Reay (1983) suggested that the point-symmetric 
features were the result of precession due to a binary nucleus with an orbital period on the order of a few days. Unfortunately, 
since then no efforts have been made in the literature to determine if NGC~5189 has a close binary nucleus. The central star has a 
[WO1] spectral type (Crowther et al. 1998) and is amongst the hottest of all central stars at 165$^{18}_{-8}$ kK (Keller et al. 2014). 
Ciardullo \& Bond (1996) monitored the central star for $\sim$9 hours and found a low-amplitude pulsation period of $11.51\pm0.05$ min that is 
typical of stars of its ilk (Ciardullo \& Bond 1996; Gonzalez-Perez et al. 1996), but noted no other variability that might otherwise indicate a companion is present.

\subsection{SAAO 1.9-m spectroscopy}
Table \ref{tab:logofobs} gives a summary of the observations with the SpCCD spectrograph on the SAAO 1.9-m. They were conducted over two separate weeks namely May 16 to 23 of 2012 and June 26 to 4 of 2012.
We use the modified julian date (MJD=JD-2400000.5) as the time of the middle of each exposure.
A slit width of 1.5" was used. Both grating 4 (1200 lines mm$^{-1}$) and 
6 (600 lines mm$^{-1}$) were used in the observing runs. A dispersion of 0.49 and 1.09 \AA/pix for the high resolution and low resolution spectra were obtained, respectively.
The wavelength coverage of grating 4 is 855 \AA\ at $\sim$ 1.09 \AA\ resolution (FWHM) and 1840 \AA\ at $\sim$ 2.4 \AA\ resolution (FWHM) for grating 6. 
The CCD was read out with a binning factor of 1$\times$2 in dispersion and spatial axes, respectively. A Copper Argon (CuAr) arc lamp was taken before and after each exposure to calibrate the wavelength scale. 
Both short ($\sim$ 240s) and long exposures ($\sim$ 2400s) were taken for each target, since we needed both unsaturated nebular emission lines and deep enough exposures to reach a S/N of $>$ 30 -- 40 in the continuum. 
The S/N was measured as an average of 4 different S/N at wavelengths $\sim$ 4220 \AA, 4755 \AA, 5290 \AA\ and 5620 \AA\ for the blue spectra and $\sim$ 5990 \AA, 6415 \AA, 6950 \AA\ and 7370 \AA\ for the red. 
The mean spectral resolution varied from 2.5 \AA\ to 2.9 \AA\ . The observations were mostly made when the seeing 
ranged between $\sim$ 1.5" to 2.5". The wavelength range was chosen according to the moon phase (blue wavelengths for grey/dark and red wavelengths for bright).

The 2D frames were trimmed and the overscan region bias level was subtracted. The L.A. Cosmic (van Dokkum 2001) \textsc{iraf} task was used to cosmic-ray-clean the images. 
Wavelength calibration was performed in the standard fashion and the mean rms achieved in the \textsc{iraf} task \textsc{identify} was $\sim$ 0.09 and $\sim$ 0.15 for high and low resolution spectra, respectively.
Each solution was applied to the 
appropriate object spectrum by adding the FITS header keywords \textsc{refspec1} and \textsc{refspec2} in the header of the object and processing them with the \textsc{iraf} task \textsc{dispcor}. The two arcs were interpolated for the final solution, in case there 
was any small shift in the arc during the exposure (e.g. due to influence of gravity on the grating causing flexure). We did not find evidence of any substantial shift in the arcs.
One-dimensional spectra were extracted using the
\textsc{apall} task. The CSPN was extracted, subtracting the immediate nebular background (4 pixels either side). However, for late [WC] types this becomes difficult due to a lack of clearly-defined nebular emission lines.

\subsection{SALT RSS spectroscopy}
We obtained 14 RSS spectra of NGC~5189 with the 
queue-scheduled SALT under programme
2013-2-RSA-005 (PI: Miszalski). Table~\ref{tab:logofobs} includes the log of
observations taken during January to March 2014. The PG2300 grating was used at a camera
articulation angle of 70 deg to cover $\lambda=$4442--5462 \AA\ which
contains several of the highest ionisation emission lines from the
CSPN of NGC~5189 (e.g. O~VI, N~V, He~II) which are formed in the wind
closer to the [WO] star than lower ionisation species would (e.g. in later
type [WC] stars), thereby improving the chances of detecting orbital
motion (see e.g. Sana et al. 2013). The 1.5 arcsec wide slit was used
at a position angle of 90 deg and resulted in a mean resolution of 2.1 \AA\ (FWHM) and
a mean dispersion of 0.32 \AA/pix. The S/N was measured as an average of 4 different S/N at wavelengths $\sim$ 4580 \AA, 4770 \AA, 5040 \AA\ and 5225 \AA\ for all spectra.
A ThAr arc lamp exposure was taken before and
after each science exposure. Basic reductions were applied using the
\textsc{pysalt}\footnote{http://pysalt.salt.ac.za} package (Crawford
et al. 2010) and cosmic ray events were cleaned using the
\textsc{lacosmic} package (van Dokkum 2001). Wavelength calibration
was performed using standard \textsc{iraf} tasks \textsc{identify},
\textsc{reidentify}, \textsc{fitcoords} and \textsc{transform} by
identifying the arc lines in each row and applying a geometric
transformation to the data frames. Both the before and after arc
exposures were used to derive the fit for the geometric transformation
and the resulting wavelength solution for all spectra had an RMS of 0.06
 $\pm$ 0.02 \AA. One-dimensional spectra were extracted using the
\textsc{apall} task, namely a main extraction of the CSPN where the
immediate nebular background (10 pixels on either side corresponding to 5.1 arcsec) was subtracted
and a second extraction of the nebula were the subtracted background 
was taken from outside the nebula (15 pixels on one side only corresponding to 7.6 arcsec) to permit measurement of nebular
radial velocities close to the CSPN position. No flux calibration was
performed as the spectra were normalised for the cross-correlation
analysis (see Figure~\ref{fig:rep2}).

\begin{figure*}
      \begin{center}
           \includegraphics[scale=0.6,angle=0]{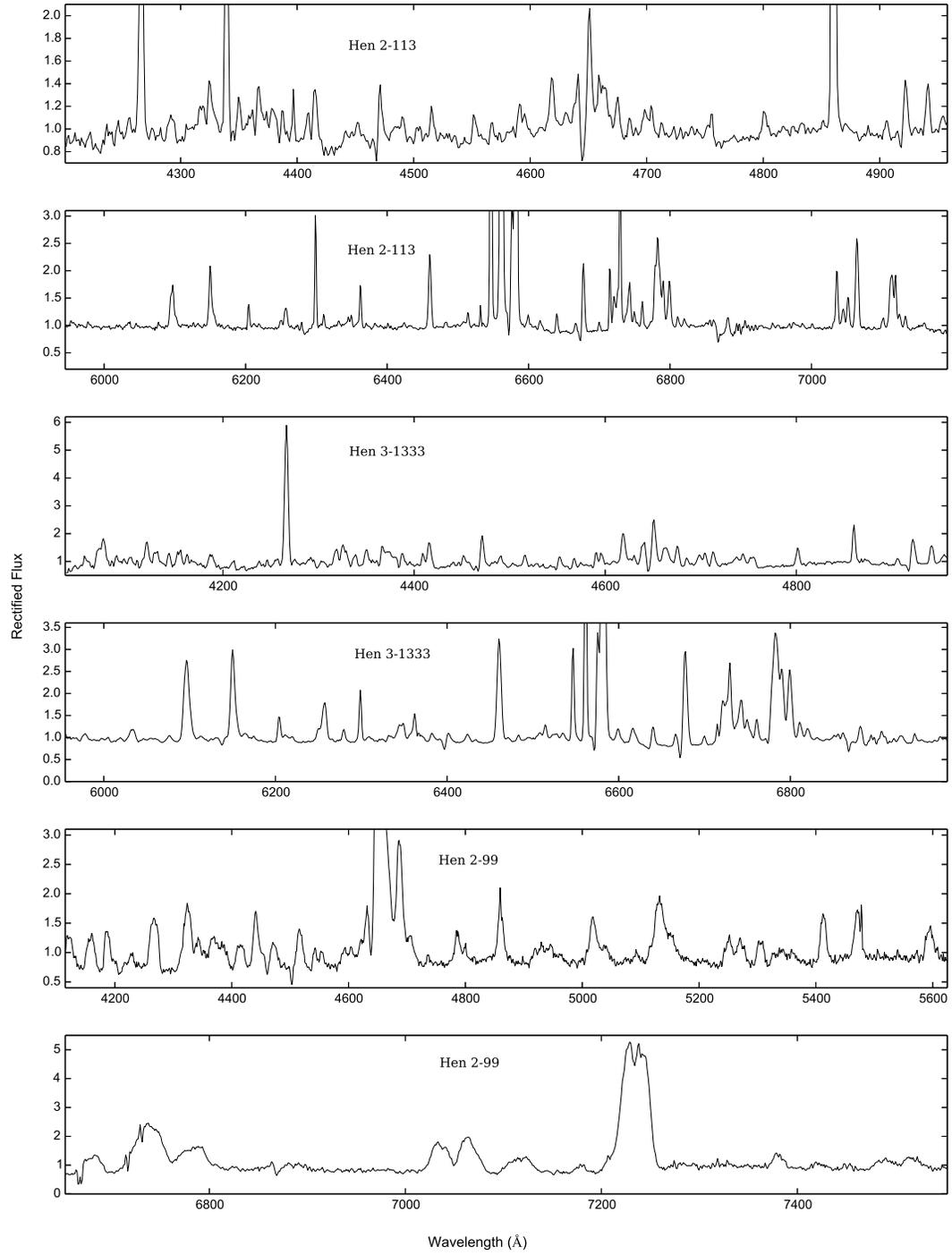}
      \end{center}
      \caption{Representative spectra of objects in our sample.}
      \label{fig:rep1}
   \end{figure*}

\begin{figure*}
      \begin{center}
           \includegraphics[scale=0.6,angle=0]{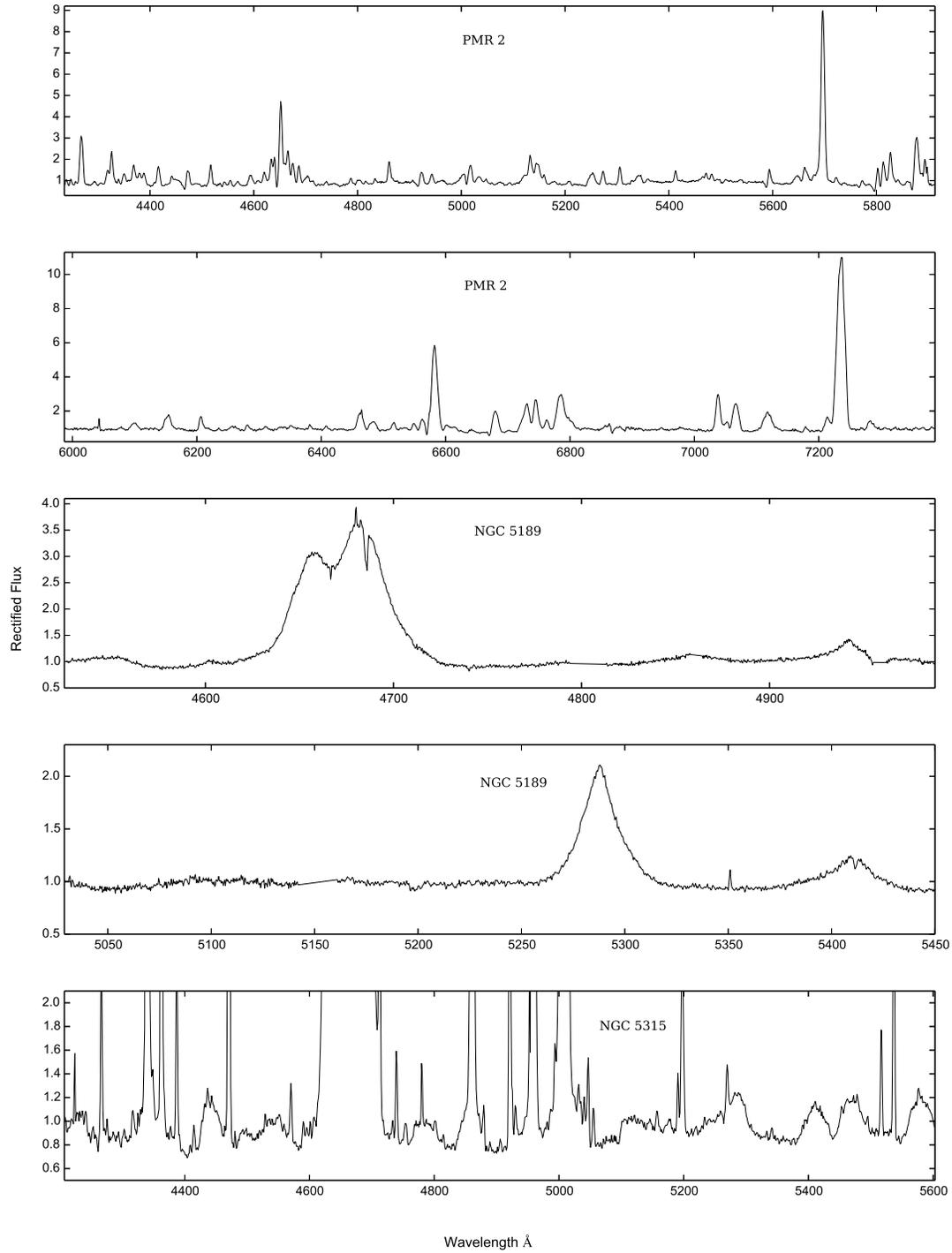}
      \end{center}
      \caption{Figure 1 (continued). Note the HeII 4686 nebular residual in NGC 5189 and the chip gaps at $\lambda$ = 4955 \AA\ to 4961 \AA\ and $\lambda$ = 5142 \AA\ 5161 \AA.}
      
      \label{fig:rep2}
   \end{figure*}

In Figures \ref{fig:rep1} and \ref{fig:rep2} we show representative spectra of each object split over the whole wavelength range. 
The strongest over-subtracted nebular lines in the spectra of NGC 5189 have been interpolated over.

\begin{table*}
   \centering
   \caption{Table showing the log of observations for the sample.}
   \label{tab:logofobs}
   \begin{tabular}{l|cccccccc}
     \hline
Object    & MJD & Date & Exp. time & Grating & $\lambda$ & Dispersion  & $\langle$ S/N $\rangle$ &  FWHM  \tabularnewline 
  & (Days) & (MM/DD/YYYY)  & (s) &  & (\AA) & (\AA\ pix$^{-1}$)  & &  (\AA) \tabularnewline \hline
Hen~2-113 & 56063.812846 & 05/16/2012 & 300 & 4 & 4260-5115 & 0.49 & 25 &  1.5 \tabularnewline
& 56063.823819 & 05/16/2012 & 900 & " & 4257-5115 & 0.49 & 50 &  1.1  \tabularnewline
& 56063.839880 & 05/16/2012 & 900 & " & 4257-5115 & 0.49 & 28 &  1.1  \tabularnewline
& 56063.869755 & 05/16/2012 & 1200 & " & 4257-5115 & 0.49 & 48 &  1.3  \tabularnewline
& 56064.122231 & 05/16/2012 & 1500 & " & 4257-5115 & 0.49 & 31 &  1.3   \tabularnewline
& 56064.801191 & 05/17/2012 & 1500 & " & 4257-5115 & 0.49 & 34 &  1.3   \tabularnewline
& 56069.996481 & 05/22/2012 & 1500 & " & 4257-5117 & 0.49 & 30 &  1.3  \tabularnewline
& 56106.823001 & 06/28/2012 & 1500 & 6 & 4025-5936 & 1.10 & 40 &  2.8   \tabularnewline
& 56108.917004 & 06/30/2012 & 1500 & " & 4055-5965 & 1.10 & 80 &  2.6   \tabularnewline
& 56108.928125 & 06/30/2012 & 1500 & " & 4036-5950 & 1.10 & 83 &  2.7   \tabularnewline
& 56109.813025 & 07/01/2012 & 1800 & " & 5722-7565 & 1.06 & 79 &  2.4   \tabularnewline
& 56111.030991 & 07/02/2012 & 1500 & " & 5722-7567 & 1.06 & 53 &  2.5   \tabularnewline
 \hline
Hen~2-99 & 56068.820887 & 05/21/2012 & 1800 & 6 & 4025-5937 & 1.10 & 43 & 2.7 \tabularnewline
& 56106.744197 & 06/28/2012 & 1800 & " & 4036-5947 & 1.10 & 45 &  2.7  \tabularnewline
& 56107.745127 & 06/29/2012 & 1800 & " & 4036-5950 & 1.10 & 25 &  2.7  \tabularnewline
& 56108.866659 & 06/30/2012 & 2400 & " & 5722-7565 & 1.06 & 29 &  2.5 \tabularnewline
& 56109.747499 & 07/01/2012 & 2400 & " & 5721-7565 & 1.06 & 23 &  2.5 \tabularnewline
& 56110.976449 & 07/02/2012 & 2400 & " & 4034-5948 & 1.10 & 38 &  2.6   \tabularnewline
& 56111.821737 & 07/03/2012 & 860 & " & 4033-5947 & 1.10 & 25 &  2.7   \tabularnewline
\hline
Hen~3-1333 & 56106.927875 & 06/28/2012 & 1200 & 6 & 4036-5947 & 1.10 & 61 &  2.6 \tabularnewline
&56108.109317 & 06/29/2012 & 1800 & " & 4032-5940 & 1.10 & 46 &  2.7 \tabularnewline
&56108.899327 & 06/30/2012 & 1800 & " & 5722-7567 & 1.06 & 34 &  2.4 \tabularnewline
&56109.131929 & 07/01/2012 & 2400 & " & 5722-7570 & 1.06 & 50 &  2.4 \tabularnewline
&56109.873401 & 07/01/2012 & 1500 & "& 5722-7572 & 1.06 & 50 &  2.4 \tabularnewline
&56109.996351& 07/01/2012 & 1500 & " & 4035-5948 & 1.10 & 64 &  2.7 \tabularnewline
&56111.057141 & 07/03/2012 & 1500 & " & 4029-5948 & 1.10 & 89 &  2.6 \tabularnewline

\hline
NGC~5315 & 56068.848701 & 05/21/2012 & 1500 & 6 & 4025-5939 & 1.10 & 30 &  2.7 \tabularnewline
& 56106.777051 & 06/28/2012 & 1500 & " & 4036-5947 & 1.10 & 24 &  2.7 \tabularnewline
& 56109.782759 & 07/01/2012 & 2400 & " & 4033-5944 & 1.10 & 24 &  2.7 \tabularnewline
& 56110.773799 & 07/02/2012 & 2400 & " & 4014-5929 & 1.10 & 23 &  2.7 \tabularnewline
\hline
PMR~2 & 56106.715497 & 06/28/2012 & 1800 & " & 4036-5948 & 1.10 & 32 &  2.6    \tabularnewline
& 56107.709237 & 06/29/2012 & 1800 & " & 4036-5948 & 1.10 & 27 &  2.6    \tabularnewline
& 56108.713127 & 06/30/2012 & 2400 & " & 5722-7565 & 1.06 & 40 &  2.5 \tabularnewline
& 56109.713509 & 07/01/2012 & 2400 & " & 5722-7565 & 1.06 & 30 &  2.4 \tabularnewline
& 56109.843839 & 07/01/2012 & 2400 & " & 5722-7565 & 1.06 & 15 &  2.6 \tabularnewline
& 56110.721211 & 07/02/2012 & 1800 & 6 & 4036-5948 & 1.10 & 37 & 2.7  \tabularnewline
& 56110.735899 & 07/02/2012 & 2400 & " & 4015-5927 & 1.10 & 36 &  2.7    \tabularnewline

\hline 
NGC~5189 &56672.581101  &  01/15/2014 & 900 &PG2300   & 4440-5462  & 0.32      &     63   &   2.1     \\
&56674.596367 &  01/17/2014  & 900 & "   &    "&  "     &     69   &   2.1                  \\
&56677.556460 &  01/20/2014  & 900 & "   &  "  &   "    &     51   &   2.1               \\
&56685.583323 &  01/28/2014  & 900 & "   &  "  &  "     &     87   &   2.1         \\
&56698.625846 &  02/10/2014  & 900 &  " &  "  &  "    &     65   &   2.1                \\   
&56700.550927 &  02/12/2014  & 900 & "   &  "  &  "     &     47   &   2.1  \\
&56701.611668 &  02/13/2014 & 900 & "   &  "  &   "    &     59   &   2.1                   \\
&56704.578300 &  02/16/2014  & 900 & "   &   " &   "    &     44   &   2.1                \\            
&56705.486876 &  02/16/2014  & 900 & "   &  "  &  "     &     61   &   2.1\\
&56705.506708 &  02/17/2014  & 900 & "   &  "  &  "     &     44   &   2.1      \\
&56706.566535  &  02/18/2014  & 900& "  &   "   & "      &     63   &   2.1    \\
&56707.556766 &  02/19/2014  & 900 & "   &  "  &  "     &     65   &   2.1   \\
&56731.589104 &  03/15/2014  & 900 & "   &  "  &   "    &     55   &   2.1                    \\
&56733.414920 &  03/16/2014  & 900 & "   &  "  &  "     &     89   &   2.1    \\

\hline

   \end{tabular}
\end{table*}

\section{Method} \label{sec:sec3}
\subsection{Cross-correlation method \& template construction}
The method used for finding radial velocity (RV) shifts in the spectra is cross-correlation as described by Foellmi et al. (2003). The cross-correlation task used 
is \textsc{xcsao} found in the \textsc{rvsao} package (Kurtz \& Mink 1998) that is executed in the \textsc{iraf} environment. \textsc{xcsao} multiplies the Fourier Transform (FT) of the object spectrum 
with the conjugate of the transform of a high S/N ratio zero-velocity template. For a detailed overview of the cross-correlation analysis, see Tonry \& Davis (1979).
This method has proved to be very useful in finding RV shifts in close massive WR binaries (Foellmi et al. 2003; Sana et al. 2013; Schnurr 2008; Schnurr et al. 2009). Despite the strong winds in [WR] CSPNe, this method is still applicable for finding binaries in these systems.

For a consistent RV monitoring programme, the cross-correlation template must have high S/N in the continuum. The object spectra need to be cross-correlated 
with this high SNR zero-velocity template to obtain accurate RV shifts.
After having applied a heliocentric correction to the radial velocities using the \textsc{bcvcorr} package in \textsc{iraf}, the template was constructed. 
The main steps involved in creating the template is outlined as follows:

\begin{enumerate}
 \item Normalise all spectra.
 \item Subtract unity from the continuum.
 \item Convert all spectra to a logarithmic wavelength scale.
 \item Use the spectrum which has the highest S/N in the continuum as first template (T1) for cross-correlation.
 \item Use the output by \textsc{xcsao} of the first cross-correlation results to shift all spectra to the same ``zero-velocity'' as the T1 template.
 \item Finally, combine all the shifted spectra to create a zero-velocity template (T2) using the \textsc{iraf} task \textsc{scombine} providing greater weights to the 
 spectra having higher S/N in the continuum.
\end{enumerate}

Once the template T2 was built, the cross-correlation is run again on the spectra, but with T2 as the template. The spectral regions which contain most of the stellar 
lines (excluding the nebular emission lines as far as possible) were used for cross-correlation because nebular lines may dilute any shifts measurable from the stellar lines.
Table~\ref{tab:ranges} gives an outline of the wavelength ranges used for cross-correlation. 
Following the Foellmi et al. (2003) method we used the bisector method to determine the absolute zero-point velocity of T2 to put all the other measurements into heliocentric velocities. Table~\ref{tab:abs_RV} summarises the strongest stellar emission lines chosen for this purpose together with their mean heliocentric velocity and the mean RV of the nebula measured separately from our observations. The online tool \textsc{vospec} (Osuna et al. 2005) was used for bisecting the chosen emission line in the template spectrum and to obtain the absolute RV of the template using the central wavelength (observed) and rest wavelength obtained from the NIST website\footnote{http://www.nist.gov/pml/data/asd.cfm}. The fraction of the emission line height ($h$) used to obtain the middle point varied for different objects. For different lines, an estimate was made to the portion which is more or less vertical i.e. $\sim$ 0.3$h$ to 0.8$h$. 
Seemingly, \textsc{vospec} was introducing a $\sim$0.5 \AA\ shift in the template, but we corrected for this in our analysis and it does not affect our results. 

\tabcolsep=0.12cm
\begin{table}
   \caption{The wavelength ranges chosen for the cross-correlations of stellar lines. The empty fields (``-'') mean either no observations were made in that particular 
wavelength range or not enough spectra were available for cross-correlation.}
\label{tab:ranges}

\begin{tabular}{lcc}
\hline
Object  &  Blue wavelength range &   Red wavelength range \\ 
 &  (\AA) &   (\AA) \\  \hline
Hen 2$-$113  & 4477-4705 \& 5050-5500    & 6064-6372 \& 6736-7050     \\ 
Hen 3$-$1333  & 4500-4720 \& 4785-5420 & 6000-6288 \& 6734-7057    \\ 
Hen 2$-$99  &  4500-5740 & 6630-7300 \& 6000-6490      \\ 
NGC 5315 & 4477-4705 \& 5050-5500 & -      \\ 
PMR 2  & 4477-5400  & 6321-6536 \& 6838-7055    \\ 
NGC 5189  & 5200-5450   & -     \\ 
\hline
\end{tabular}

\end{table}

\begin{table*}
\centering
\caption{The weighted mean radial velocities for stellar and nebular emission features of each object. The nebular lines used are H${\alpha}$, H${\beta}$, H${\delta}$, H${\gamma}$, NII-6583 and OIII-4959.}
\label{tab:abs_RV}
\begin{tabular}{l|ccr||cr}
\hline
Object  & Stellar   & $\lambda_{rest}$& $\langle$RV$_\mathrm{stellar}$$\rangle$   &  Nebular & $\langle$ RV$_\mathrm{neb}$$\rangle$  \\ 
  &    &  (\AA) &  (km s$^{-1}$)   &    &  (km s$^{-1}$)  \\  \hline
Hen 2-113 & CII&  4267.26 & $-$60$\pm$4  &H${\alpha}$, H${\beta}$, H${\delta}$, H${\gamma}$, NII & $-$63$\pm$2 \\
Hen 3-1333 & CIII & 5695.19 & $-$43$\pm$4 & H${\alpha}$, H${\beta}$, H${\gamma}$, NII & $-$66$\pm$10\\
Hen 2-99 & CIII& 5695.19 & $-$56$\pm$12  & H${\alpha}$, H${\beta}$, NII & $-$95$\pm$13  \\
NGC 5315 & HeI &  4471.47&84$\pm$15  & H${\alpha}$, H${\beta}$, H${\gamma}$, NII, OIII & $-$24$\pm$8 \\
PMR~2 & CII &  4267.18& 4$\pm$5 & H${\alpha}$, H${\beta}$, NII & $-$66$\pm$23  \\
NGC 5189 & OVI& 5290.65 & 23$\pm$40  & $H_{\beta}$ & $-$8$\pm$4 \\
\hline
\end{tabular}
\end{table*}

\subsection{Construction of RV time-series}

The main outputs by \textsc{xcsao} are the RV shift in km s$^{-1}$ of the object 
spectrum relative to the T2 template together with the height of the cross-correlation 
peak and the associated error in km s$^{-1}$. In summary, the velocity obtained by the bisector represents the mean stellar velocity and the RVs are computed using:

\[
 v_i = v_{xcsao} + v_{bisector} + v_{hcv}
\]
Where $v_{xcsao}$ is the RV output by \textsc{xcsao}, $v_{bisector}$ is the stellar mean RV calculated using the bisector method and 
$v_{hcv}$ is the heliocentric velocity correction.

The error in the individual RVs ($v_i$) was computed in quadrature using the equation:

\[
 \triangle v_i = \sqrt{\triangle v_{xcsao}^2 + \triangle v_{bisector}^2}
\]

Where $\triangle v_{bisector}$ is the error measurement for the bisector method and $\triangle v_{xcsao}$ is the error output by \textsc{xcsao} as described by Kurtz \& Mink (1998). A peak is selected by fitting a smooth curve with accurate values of $\delta$ 
(the central value), $h$ (the height of the central peak) and $w$ (the FWHM of the peak). Kurtz \& Mink (1998) assumed a sinusoidal noise profile, with 
the halfwidth of the sinusoid equal to the halfwidth of the correlation peak. The mean error output by \textsc{xcsao} is a single velocity measurement and is given by:
 
\[
 error = \frac{3}{8} \frac{w}{(1+r)} 
 \]
Where $w$ is the FWHM of the correlation peak and $r$ is defined in Tonry \& Davis (1979), as the the ratio of the height of the true peak to the average peak.

\section{Results}\label{sec:sec4}

\subsection{RV time series}
Tables \ref{tab:appendix} and \ref{tab:appendix1} list the resultant RV time-series for each object. Figures \ref{fig:rvs_all_1} and \ref{fig:rvs_all_2} display the data graphically. We emphasise that since the [WR] emission lines are formed in the wind, the absolute values of the velocities may not be meaningful (especially when compared against the nebular RVs). To quantify the variability level observed in each object, we conducted a $\chi^2$ analysis of the RV time-series based on a null hypothesis which is discussed in detail by Trumpler \& Weaver (1953). These authors recommend a standard $\chi^{2}$ computation, from the reduced $\chi_{n-1}^{2}$ 
(where n is the number of RV measurements and n-1 is the number of degrees of freedom) which leads to an estimate of the probability that the star's RV is variable. 
In summary, let's say the hypothesis to be tested is $H:$ \textit{The semi-amplitude ``A'' of a radial velocity variation equals zero.}, i.e. there is no statistical difference
between the observed values ($x_j$) and the expected values ($\bar x$).
Then alternatively, the hypothesis is that $A > 0$. The test for $H$ is such that:

   \begin{figure*}
      \begin{center}
        \includegraphics[scale=0.7]{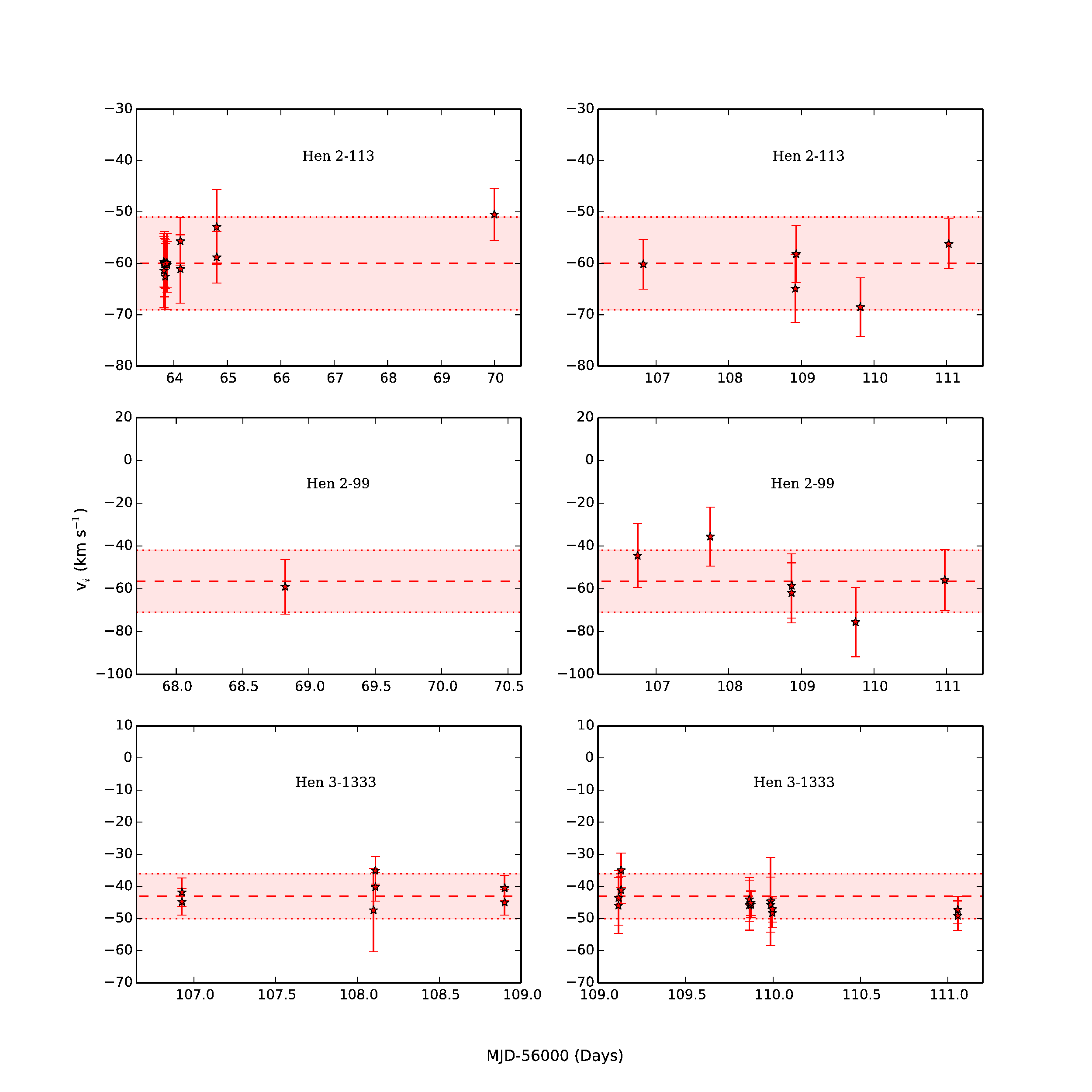}
      \end{center}
      \caption{Measured RV time-series for the nuclei of Hen 2-113, Hen 2-99 and Hen 3-1333.}
      \label{fig:rvs_all_1}
   \end{figure*}

   \begin{figure*}
      \begin{center}
        \includegraphics[scale=0.7]{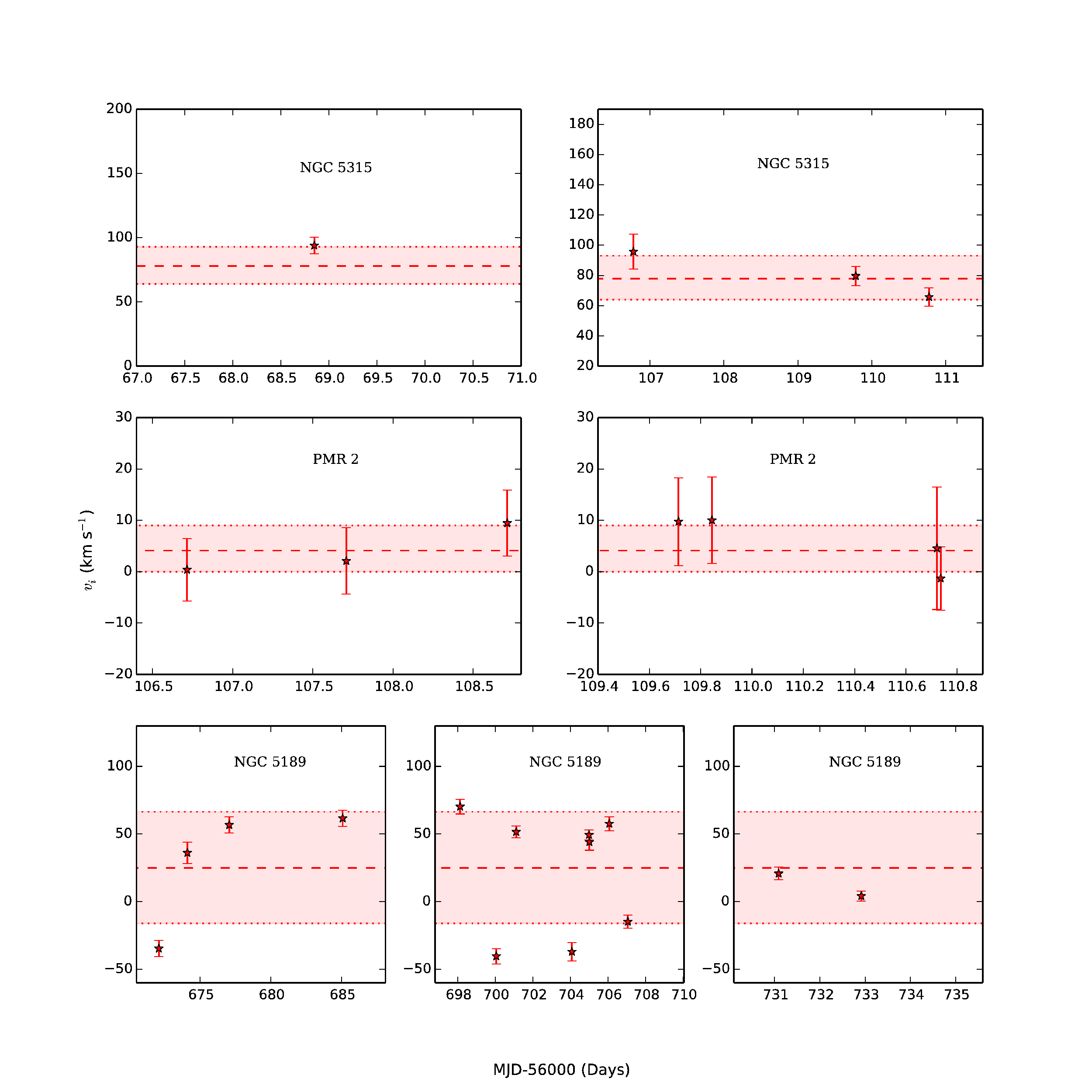}
      \end{center}
\caption{Measured RV time-series for the nuclei of NGC 5315, PMR 2 and NGC 5189.}
      \label{fig:rvs_all_2}
   \end{figure*}

      \begin{enumerate}
\item{$P_{I}$ is the probability of \textit{rejecting} $H$ when $H$ is true (i.e, saying $A > 0$, when $A = 0$). Now, if we set $\alpha$ = 0.05 arbitrarily, then in this case, $P_{I}$ $\leq$ $\alpha$.}
\item{$P_{II}$ is the probability of \textit{accepting} $H$ when $H$ is false (i.e, saying $A = 0$, when $A > 0$). In this case $P_{II}$ must be a minimum, 
no matter which orbit. So, maximizing $\beta$ (where $\beta$ = 1 - $P_{II}$, is the probability that the test will reject $H$ when $H$ is false) in the vicinity 
of $A = 0$.}
\end{enumerate}

The compromise consists of allowing the value of $\beta$ to be at least as large as $\alpha$. Hence, the criterion for rejection is: reject the hypothesis 
that $A = 0$ whenever:

\[
 \sum_{j=1}^n \frac{(x_{j}-\bar x)^2}{\sigma^2_i} \geq \chi^2_{\alpha, n-1}
\]

where  
\begin{itemize}
   \item $\chi_{\alpha, n-1}^2$ is the standard $\chi^2$ value for $n-1$ degrees of freedom and probability, $P$ $=$ $\alpha$. 
\item $\bar x$ is the stellar mean (Expected value).
\item $x_{j}$ is the individual values for the stellar RVs.
\item $\sigma_i$ the individual errors in the RVs.
\end{itemize}

The null hypothesis simply states that there is no statistical difference between the observed value (data) and the expected value which in our case is the stellar mean. 
The main point is to either accept or reject the null hypothesis with a given probability based on a critical value ($\alpha$ = 0.05). If the value of the observed $\chi^2$ is less
than $\chi^2_{0.05, n-1}$, we will accept the null hypothesis with a probability of 95\%. This means that we are 95\% sure that the variations in the RV are random.
Contrarily, if the observed $\chi^2$ is greater than $\chi^2_{0.05, n-1}$ then we will reject the null hypothesis with a probability of 95\%, meaning the RV is highly variable. The convention we will use here is looking at the probability of accepting the null hypothesis.

The above variability test was run on the 6 objects, assuming the stellar mean is the expected mean and the results are shown in Table \ref{tab:chi2}. 
For each of them, the $\chi_{obs}^{2}$ (observed Chi-squared) was computed and the probability that the object is variable $P$($\chi_{obs}^{2}$ $\geq$ $\chi_{n-1}^{2}$) was obtained. For some stars, the RV variability was so large that it becomes obvious with only a few observations to decide whether it is a binary or not. However, in most cases we lack sufficient observations to properly constrain the amount of variability in the RV. In cases where small random fluctuations in the RV from the mean occur, which we would expect due to instrumental uncertainties or wind variability, we would identify the RV as being constant. However, a highly variable RV from the mean would be one having large enough RV variations that cannot be assigned solely to random variations. In the case of a close binary CSPN we require a statistically significant periodic signature in the RV time-series to be present.

\begin{table}
\centering
\caption{Results of the variability test carried on individual objects. Column 1 shows the object name, the 2$^{nd}$ column is the number of RV measurements 
(degrees of freedom$+$1), the 3$^{rd}$ column is the mean RV, the 4$^{th}$ column is the standard deviation in the RV, the 5$^{th}$ column is
the observed $\chi^{2}$ and
column 6 shows the probability of accepting the null hypothesis based on the $\chi^{2}$ value.}
\label{tab:chi2}
\begin{tabular}{l@{\hskip 0.2in}|r@{\hskip 0.2in}r@{\hskip 0.2in}r@{\hskip 0.2in}r@{\hskip 0.2in}r}
\hline
Object  & $n$ & $\langle$RV$\rangle$  & $\sigma$ & $\chi_\mathrm{obs}^{2}$ & P$_\mathrm{null}$\\
  &   &  (km s$^{-1}$) &(km s$^{-1}$) & &  \\ \hline
Hen2-113 & 18 & $-$59.5 & 4.1 & 8.8 &   94\%  \\
Hen3-1333 & 15 & $-$44.1 & 4.4 & 0.2 &   100\% \\
Hen2-99 & 8 & $-$55.7 & 12.0 & 4.5 &   72\% \\
NGC 5315 & 4 & 87.3& 14.0 & 15.3 &  1\% \\
PMR~2 & 7 & 4.0 & 4.8 & 3.0 &  80\% \\
NGC 5189 & 14 & 23.2 & 40.1 & 690 &  0\%\\
\hline
\end{tabular}
\end{table}%

The objects observed with the SAAO 1.9-m are not suffiently variable to prove a binary companion is present at this time, 
however two objects stand out from the rest. Hen 2-99 has stellar RV shifts with a peak-to-peak value of the order 30 km s$^{-1}$ 
from the mean and a hint of periodicity is seen with a peak-to-peak of $\sim$ 53 km s$^{-1}$. A sinusoid fits reasonably well 
with the RV of Hen 2-99 with a period of 5.3 d. More data is needed to confirm this period. Similarly, NGC 5315 shows much higher 
RV variability ($P_\mathrm{null}=1$ \%), but we have only 4 epochs for this object. The non-periodic and relatively small amplitude 
RV shifts in the remaining objects that we see are most probably due to stochastic wind variability (e.g. Grosdidier et al. 2000, 2001). 

NGC 5189 is the only object in the sample that shows very high RV variability with a peak-to-peak velocity of the order of 120 
km s $^{-1}$ (see Fig. ~\ref{fig:rvs_all_2}).
From the $\chi^2$ variability analysis in Table~\ref{tab:chi2} and Fig. \ref{fig:rvs_all_2} it is quite clear that it 
is the only object which shows 100\% variability. The following section further analyses this unique dataset.

\section{The close binary CS of NGC 5189}\label{sec:sec5}

\subsection{Orbital period analysis and mass function}
\begin{figure*}
      \begin{center}
         \includegraphics[height=126mm,width=125mm,scale=0.55,angle=0]{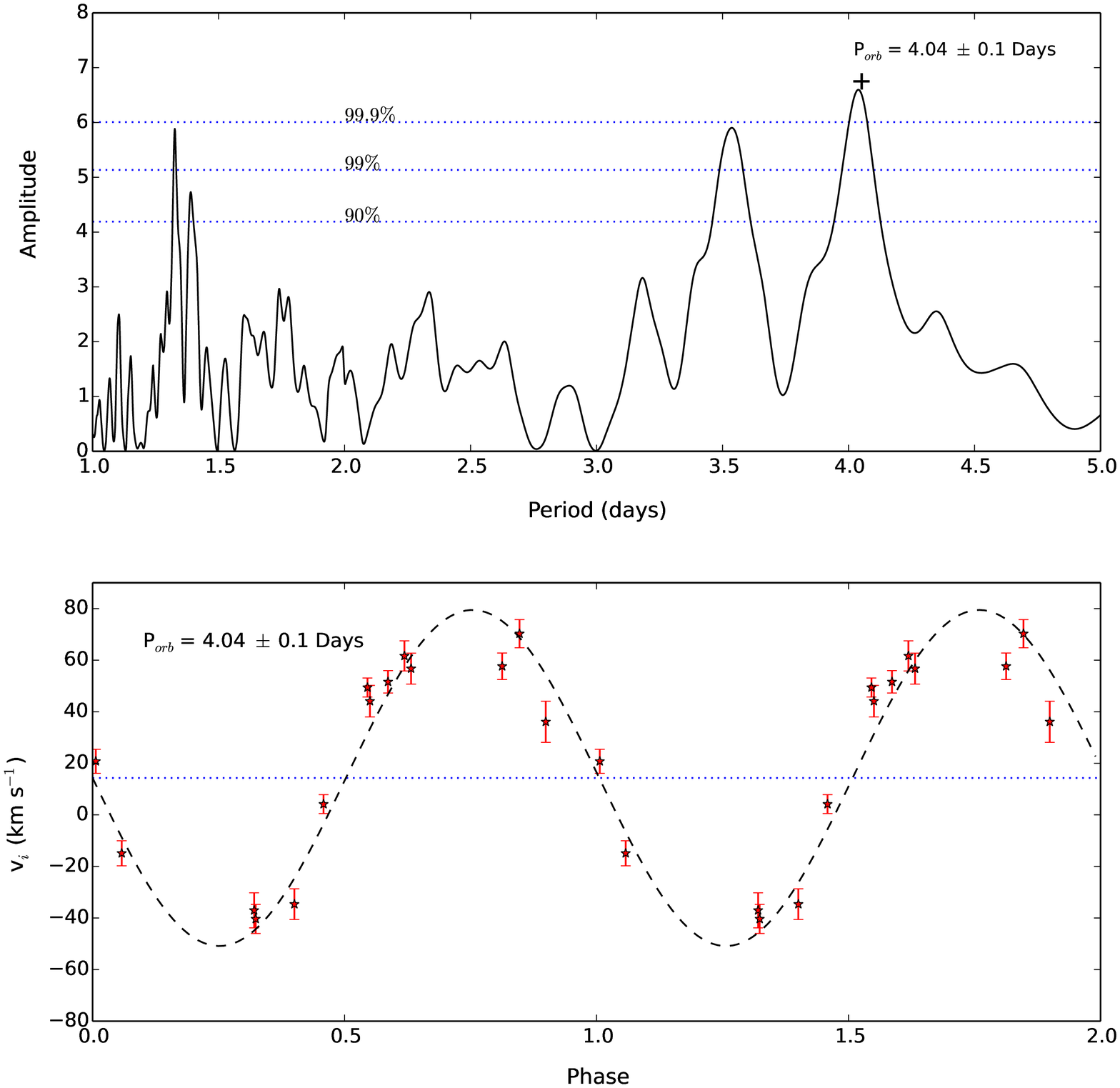} \\ \vspace{-5.mm}
         \includegraphics[height=102mm,width=125mm,scale=0.55,angle=0]{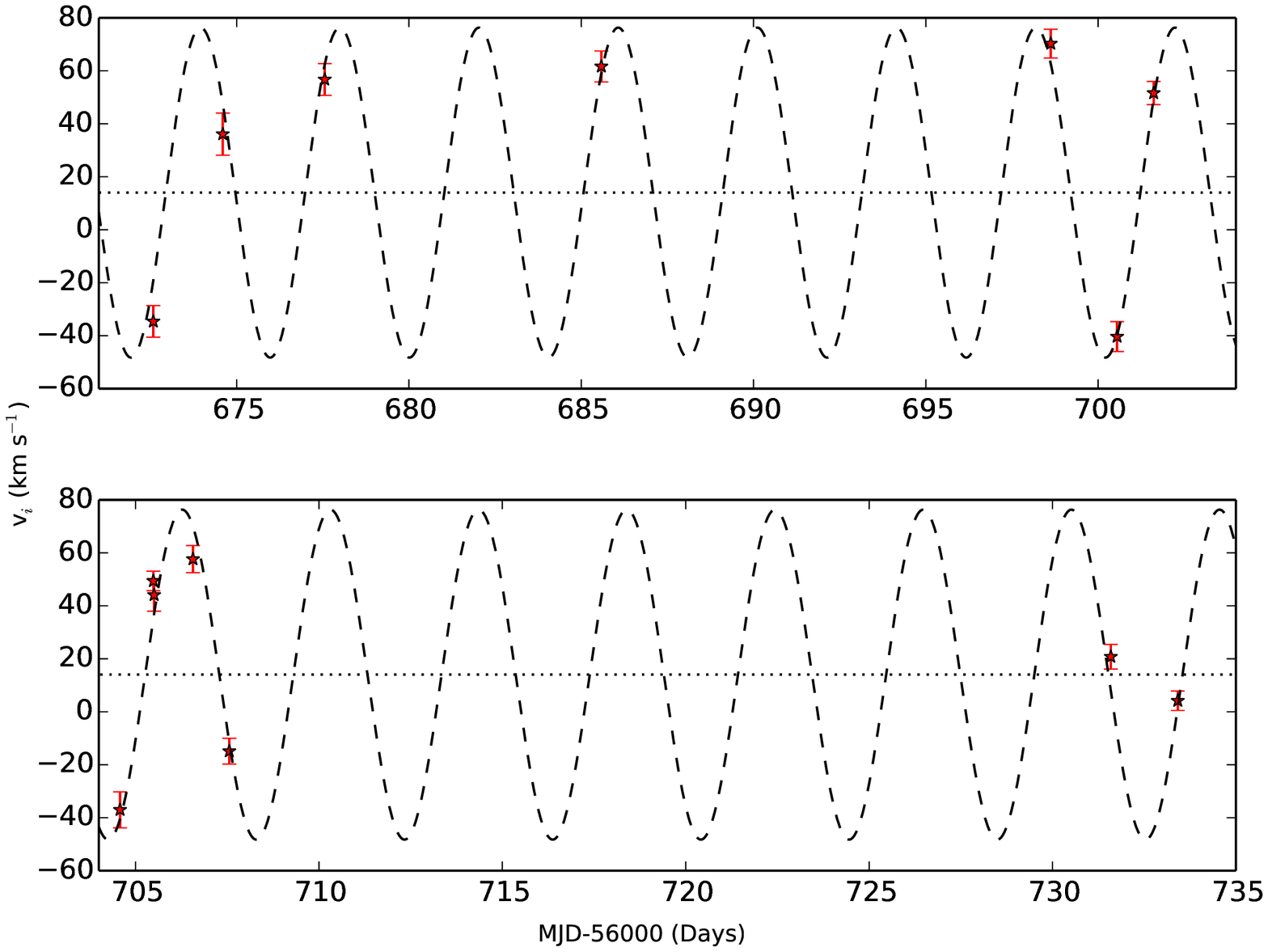}
      \end{center}
      \caption{ \textbf{Top}: The Lomb-Scargle periodogram showing the significance levels of 90\%, 99\% and 99.9\%. \textbf{Middle}: The radial velocity folded on the 4.04 day period fitted 
      with a sinusoidal model. \textbf{Bottom}: Plot of the radial velocities folded on the 4.04 day period fitted with a sinusoid.}
      \label{fig:lslc}
   \end{figure*}

Since our data is not evenly sampled, we used the Lomb-Scargle (Lomb 1976; Scargle 1982) method to search for an orbital period. 
The Lomb-Scargle method is mainly based on a least-squares fit of sinusoids (Press \& Rybicki 1989). Figure~\ref{fig:lslc} shows 
four significant peaks in the periodogram at periods 4.04 days, 3.54 days, 1.32 days and 0.80 days. 
However, we find that the 3.54 day, 1.32 day and 0.80 day peaks are most likely to be aliases, as they do not persist when 
we ran the Lomb-Scargle for a second time after extracting the 4.04 day period. Moreover,
to test the validity of the periods, we further analysed the data according to the method described by Tanner (1948). Assuming the actual period is 4.04 d, both the 1.32 d and the 0.8 d periods satisfy Equation 2 in Tanner (1948). Based on the relatively high value $\chi^2$ of 8.3 for the 3.54 day period sinusoid fit, we further reject this period. This leaves us with the best period being 4.04 d which has the
lowest $\chi^2$ of 3.0 out of all periods.

A significance test was also carried out on the periods found using
a Monte-Carlo simulation, where a random number generator was used to scramble the original radial velocities within one standard 
deviation, creating 10000 randomised RV curves of the same structure. A Lomb-Scargle analysis was then performed on the randomised 
RV curves and the highest powers were recorded for each of them. The powers were then sorted in ascending order after which the 
90\%, 99\% and 99.9\% significance levels were obtained (displayed as horizontal lines in Fig. \ref{fig:lslc}). The only period 
which was found to be above the 99.9\% level and acceptable as per the sinusoid model fit was the 4.04 day.
   
\begin{figure*}
      \begin{center}
         \includegraphics[scale=0.7,angle=0]{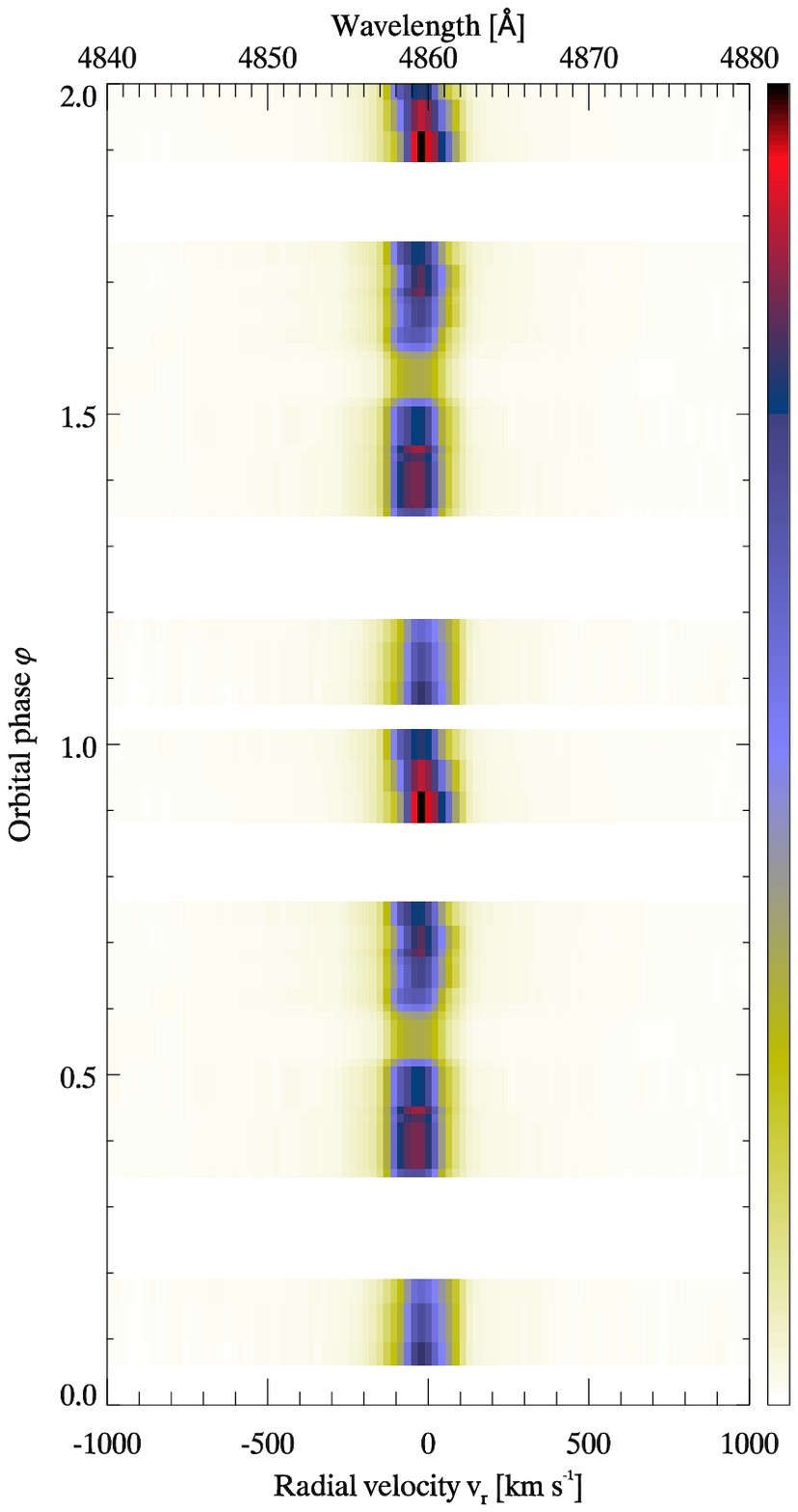}
         \includegraphics[scale=0.7,angle=0]{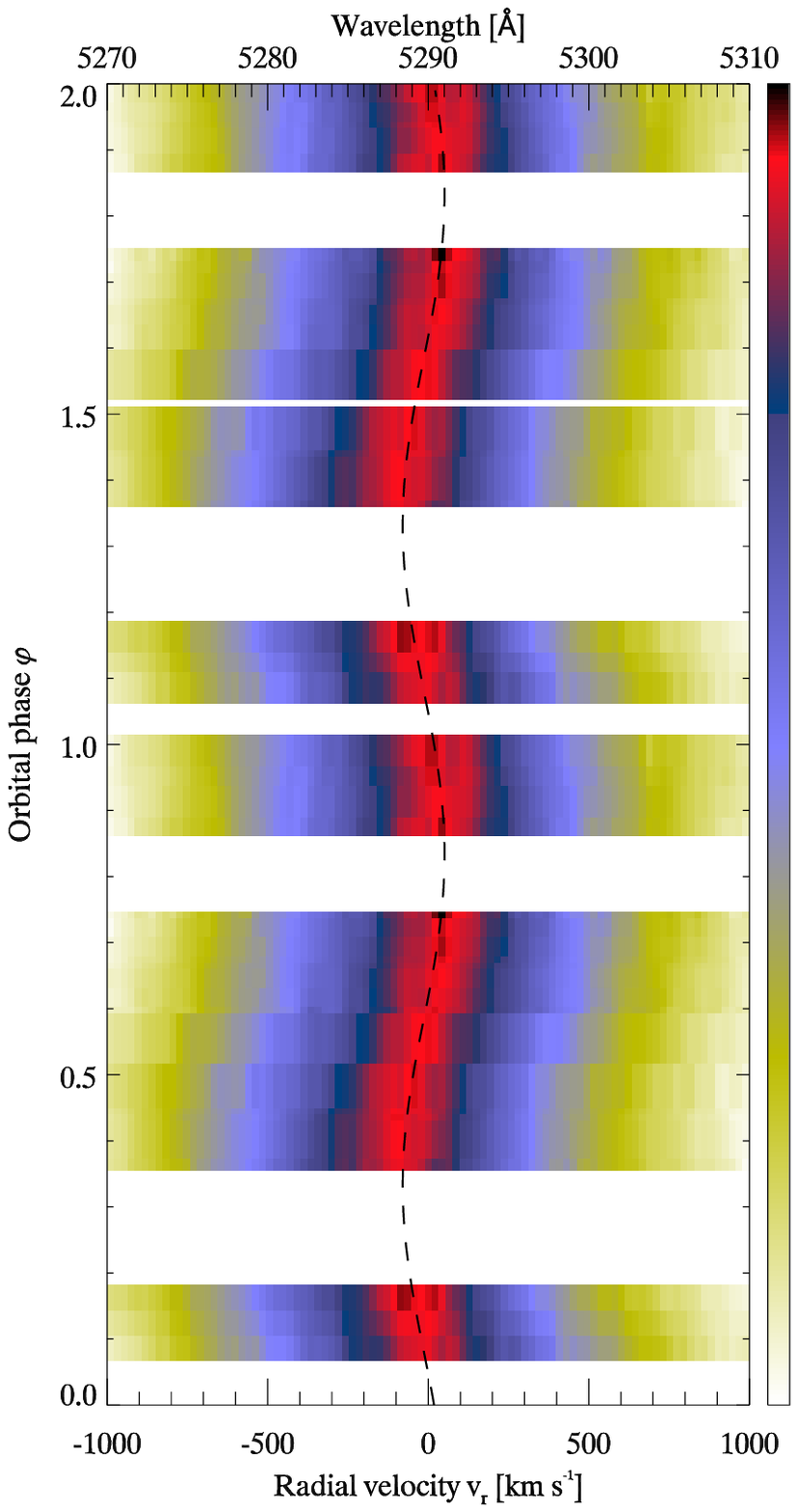}\\
      \end{center}
      \caption{Left: Trail diagram of the nebular H$\beta$ line. Right: Trail diagram of the stellar OVI-5290 emission line. The dashed line shows the sinusoidal fit to the RV curve in phase space.}
      \label{fig:trail}
   \end{figure*}

   The second panel in Figure~\ref{fig:lslc} is a sinusoid model determined from fitting the data folded on the 4.04 day period in phase space. We used the \textsc{mpfit} and \textsc{mpfitfun} \textsc{idl} codes designed to perform a Levenberg-Marquardt least-squares fit of a user supplied model to a function (Markwardt 2012). The main fitting was done by \textsc{mpfit} after specifying a set of initial parameters. Given the data and their uncertainties, \textsc{mpfitfun} computes the best set of model parameters which match the data and returns them in an array. In time space we determined a 4.04 day period fit of the form $A$ $\sin(2\pi f-\phi)$ $+$ $B$ to the data where $A$ denotes the amplitude, f is the frequency fixed at 0.247 day$^{-1}$, $\phi$ is the phase and $B$ is the mean, with $A=62.3$ km s$^{-1}$, $\phi=0.43$ and $B=14.0$ km s$^{-1}$. This corresponds to an ephemeris determined from the minimum radial velocity at MJD (min RV) = (2456700.24 $\pm$ 0.02) + (4.04 $\pm$ 0.1)$E$. The error in the orbital period was determined by fitting a gaussian with a mean of 4.04 d and a standard deviation of 0.1 d to the peak period in the periodogram. The bottom two panels of Fig. \ref{fig:lslc} shows this fit along with the RV time-series data showing the reasonable agreement to our observations.  
   Figure \ref{fig:trail} gives another graphical representation of the periodic motion in the RV time-series data of NGC~5189. The trail diagrams are plotted over 2 phases using the motion of the OVI-5290 stellar emission line and, for comparison, the static $H\beta$ nebular emission line. 

In the case of close-binaries with a good sinusoidal fit, it is fair to assume a low (e$\ll$1) or zero eccentricity. In this case
the mass function of two stars of masses $m_1$ and $m_2$, where only one of the stars is visible in the spectrum (single-line spectroscopic binaries) is described by:

\[
f(m_1,m_2)=\frac{m^3_{2}}{(m_{1}+m_{2})^2} \sin^3 \textit{ i}=\frac{P}{2 \pi \textit{G}} v^3_{1}
\]
where $P$ is the orbital period ($4.04\pm0.1$ d), $i$ is the inclination of the orbital plane to the line of sight, and $v_1$ is the semi-amplitude of the RV curve ($v_1$ = 62.3 $\pm$ 1.3 km s$^{-1}$). 

Table \ref{tab:masses} lists companion masses $m_2$ calculated for a range of inclination values and an adopted $m_1$ mass of 0.596 $M_\odot$ for the [WO1] primary. The value of $m_1$ was calculated as an average of the masses of Miller Bertolami \& Althaus (2006) evolutionary tracks either side of the location of NGC~5189 in figure 13 of Keller et al. (2014). There are several caveats associated with this assumption. Firstly, the mass estimates derived from interpolating between evolutionary tracks are distance dependent and distances to PNe are notoriously uncertain. Keller et al. (2014) adopted a distance of 0.55 kpc to NGC~5189 (Stanghellini et al. 2008), but the distance may be as large as $1.44\pm0.27$ kpc based on extinction and kinematic distance estimates (Frew 2008). The luminosity would therefore increase from $\sim2.73$ $L_\odot$ (Keller et al. 2014) to $\sim3.86$ $L_\odot$ (Frew 2008), resulting in a mass closer to $\sim$0.62 $M_\odot$. However, it is not necessarily a matter of adopting this mass as this method of estimating CSPNe masses does not agree well with spectroscopic mass estimates for CSPNe (see section 4.2.3 of Moe \& De Marco 2006 and ref. therein), underlying the intrinsic unreliability of such model-dependent methods. Finally, the evolutionary tracks assume single star evolution which may not apply to a post-CE binary such as the nucleus of NGC~5189.

In summary, these caveats make it difficult to choose any `best' value for $m_1$. As the orbital inclination of the nebula is currently unconstrained (Sabin et al. 2012), it is only possible in this work to give an indication of plausible companion masses. The unknown inclination introduces a much larger uncertainty in $m_2$ than the small uncertainty in $m_1$. Therefore, for the purposes of this work, it is reasonable to retain the adopted value of $m_1=0.596$ $M_\odot$ that is essentially the same as the standard mass assumed for CSPNe of 0.6 $M_\odot$ (e.g. Crowther et al. 2006; Crowther 2008). Even if a slightly more massive $m_1$ were adopted, it would only result in a small increase in $m_2$.

\begin{table}
\centering
\caption{Companion masses ($m_2$) for NGC~5189 for a variety of orbital inclination angles.}
\label{tab:masses}
\begin{tabular}{cccc}
\hline
Inclination  & $m_2$  & $\triangle m_2$  \\
($^\circ$) & ($M_{\odot}$) & ($M_\odot$) \\
\hline
30 & 1.56 & 0.03 \\
35 & 1.21 & 0.02 \\
40 & 0.99 & 0.02 \\
45 & 0.84 & 0.01 \\
50 & 0.74 & 0.01 \\
55 & 0.67 & 0.01 \\
60 & 0.61 & 0.01 \\
65 & 0.57 & 0.01 \\
70 & 0.54 & 0.01 \\
75 & 0.52 & 0.01 \\
80 & 0.51 & 0.01 \\
85 & 0.50 & 0.01 \\
90 & 0.50 & 0.01 \\
\hline

\end{tabular}
\end{table}
\subsection{Nature of the companion}
We give a summary of the properties of the CS of NGC~5189 in Table \ref{tab:propertiesNGC5189} and Table~\ref{tab:orbparams} gives the orbital parameters derived in this work. The mass function puts a lower limit on $m_2$ (0.5 M$_\odot$ $\pm$ 0.01) which is reached only if the system is edge-on with $i=90^\circ$. For inclinations less than 40 degrees we find improbably high values of $m_2$, suggesting these inclinations can be discarded. At other inclinations the possible companions are a main sequence star or a WD. It may be possible for an orbital period of 4.04 d that a main sequence companion could produce an irradiation effect, especially given the 165 kK temperature of the [WO1] component. This effect could be up to 0.1-0.2 mag (De Marco et al. 2008), but this has not yet been observed (e.g. Ciardullo \& Bond 1996). There is also no evidence to suggest a cool companion is present since the intrinsic $(V-I)$ colour of $-0.32\pm0.08$ is typical of hot blue CSPNe ($(V-I)_0\sim-0.4$, Ciardullo et al. 1999). The intrinsic faintness of the primary, namely $M_V=+2.7$ to $+4.8$ mag for $d=1.44$ kpc (Frew 2008) or 0.55 kpc (Stanghellini et al. 2008), respectively, also means that it is unlikely that a cool companion is hiding in the glare of the primary. Although the orbital inclination is currently unconstrained (Sabin et al. 2012), we can estimate an intermediate inclination of 45 degrees based on the apparent nebular morphology, i.e. the relatively symmetrical inner nebula with a pinched waist (see Sect. \ref{sec:morph})) and that the collimated outflows are neither completely inside or outside the nebula. The companion is therefore likely to be a slightly more massive WD of $\sim$0.9 $M_\odot$, comparable to the companion discovered in Fleming~1 (Boffin et al. 2012).

\renewcommand{\tabcolsep}{0.2cm}
\begin{table}
   \centering
      \caption{Properties of NGC~5189 and the [WO1] primary.}
   \begin{tabular}{lrl} \\
   \hline
      PN G & 307.2$-$03.4 &\\
      $c(H\beta)$ & $0.47\pm0.08$ & Garcia-Rojas et al. (2012)\\
      $V$ (mag) & 14.53 & Ciardullo et al. (1999)\\
      $I$ (mag) & 14.35: & Ciardullo et al. (1999)\\
      $(V-I)_0$ & $-$0.32$\pm$0.08 & This work \\
      Type & [WO1] & Crowther et al. (1998)\\
      $v_\infty$ (km s$^{-1}$) & 2500$\pm$250 & Keller et al. (2014)\\
      $T_\mathrm{eff}$ (kK) & 165$^{+18}_{-8}$ & Keller et al. (2014)\\
      $m_1$ (M$_\odot$) & 0.596 & Keller et al. (2014)\\
      \hline

   \end{tabular}

   \label{tab:propertiesNGC5189}
\end{table}

\renewcommand{\tabcolsep}{1cm}
\begin{table}
\centering
\caption{Orbital parameters of NGC 5189.}
\label{tab:orb_prop}
\begin{tabular}{lr}
\hline
$P_{orb}(d)$ & 4.04 $\pm$ 0.1 \\
$e$ & 0 (fixed) \\
$K$ (km s$^{-1}$) & 62.3 $\pm$ 1.3\\
$\gamma$ (km s$^{-1}$) & 14 $\pm$ 1\\
$m_2$ (M$_\odot$) & $\ge0.5$ \\
$T_0$ (MJD) & 56699.74$\pm$0.02\\
\hline
\end{tabular}
\label{tab:orbparams}
\end{table}

\begin{figure}
      \begin{center}
         \includegraphics[scale=0.35,angle=270]{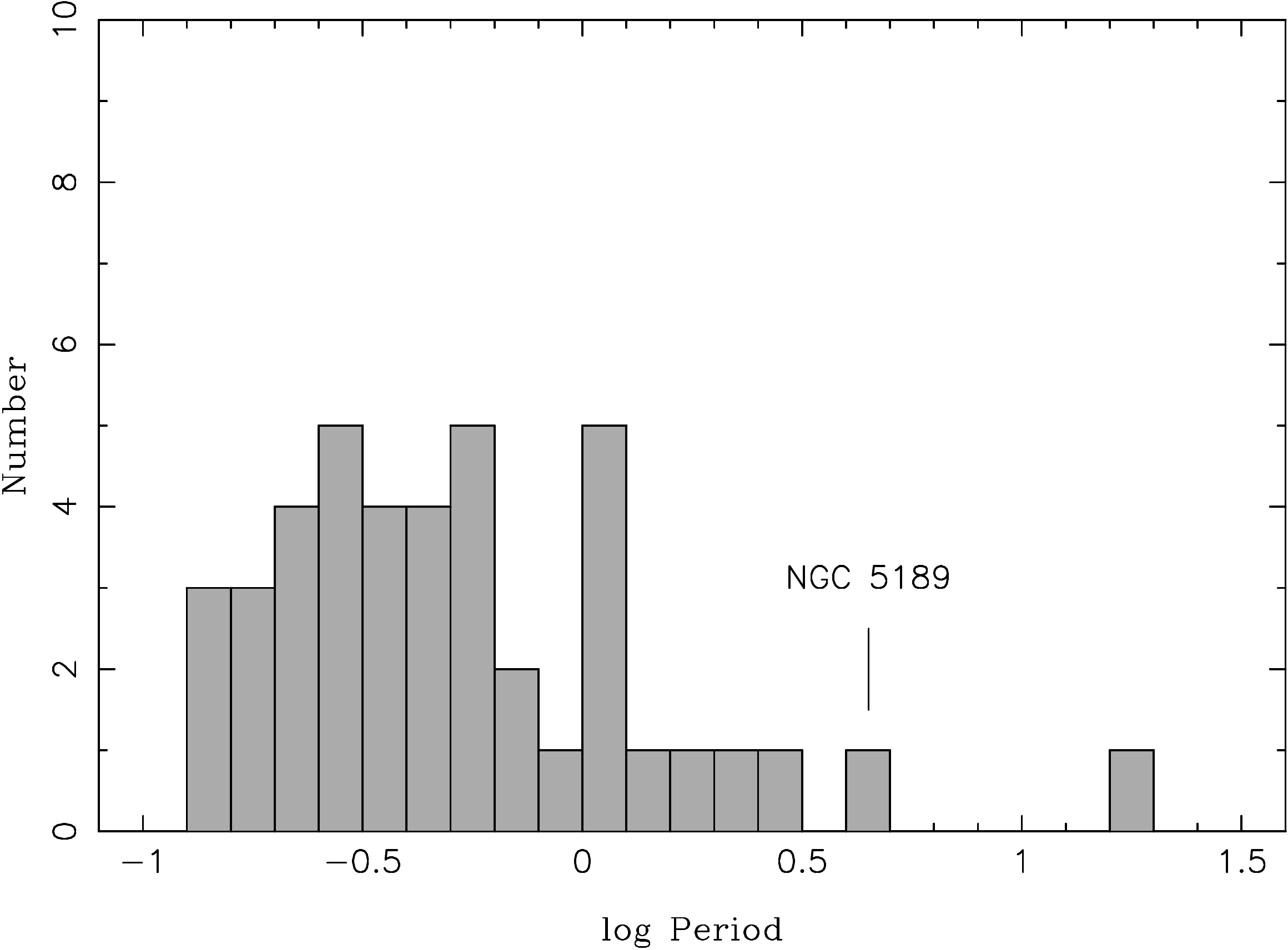}
      \end{center}
      \caption{Orbital period distribution of published close binary CSPNe (see Appendix \ref{tab:closebinaries}).}
      \label{fig:pdist}
   \end{figure}

The 4.04 d orbital period of NGC~5189 is one of the longest currently known amongst post-CE CSPNe (Fig. \ref{fig:pdist}) and falls into the zone with periods $\ga$1 day where there is a substantial deficit of post-CE binaries compared to CE population synthesis models (Rebassa-Mansergas et al. 2008; Miszalski et al. 2009a; Davis et al. 2010; Nebot G\'omez-Mor\'an et al. 2011). There are two possible interpretations for the position of NGC~5189 in the post-CE orbital period distribution. One is that the longer period is a one-off coincidence as in NGC~2346 and this would be in line with the observed deficit. The other possibility is that NGC~5189 reflects the tip of an iceberg where most [WR] close binaries would have similarly long orbital periods (or longer). This could be explained by the stronger and more extended wind of the [WR] component that could interact with the companion (e.g. via wind Roche-lobe overflow, Mohamed \& Podsiadlowski 2007, 2011). If the interaction were to facilitate the CE phase at greater orbital separations than in the average post-CE CSPN (WD and main sequence), then the resultant orbital period might be longer than the average. At present with only two [WR]  post-CE CSPNe known, it is difficult to decide which of these two possibilities is correct.

\subsection{The post-CE nebular morphology}
\label{sec:morph}
The main reason for including NGC~5189 in our study was that it contains all the hallmarks of post-CE nebulae as first outlined 
by Miszalski et al. (2009b). These hallmarks include low-ionisation structures (LIS; see e.g. Gon{\c c}alves et al. 2001), collimated outflows 
or jets and bipolar nebulae (see also Miszalski et al. 2011b and Miszalski 2012). Additional recent discoveries of post-CE nuclei in other PNe 
have further reinforced these hallmarks (Corradi et al. 2011; Miszalski et al. 2011a,b,c; Boffin et al. 2012; Jones et al. 2014). The abundance of 
LIS filaments in NGC~5189 (e.g. Sabin et al. 2012), the majority of which point to the central star, is exactly what we see in NGC~6326 (Miszalski et al. 2011b) 
and similarly high levels of LIS are also seen in NGC~6778 (Miszalski et al. 2011b; Guerrero \& Miranda 2012) and Hen~2-11 (Jones et al. 2014). 
Secondly, Sabin et al. (2012) demonstrated that the outermost LIS of NGC~5189 have the most extreme velocities of the nebula. 
Combined with the central $S$-shaped feature, these outermost LIS are symptomatic of a precessing outflow (e.g. Lopez et al. 1993) that is best explained by jets being launched from a precessing 
accretion disk around a companion (Cliffe et al. 1995; Raga et al. 2009; Boffin et al. 2012). Furthermore, leaving the LIS aside, the intrinsic morphology of 
NGC~5189 appears to be a bipolar outflow best seen in the [O~III] emission line. It is pinched at the waist at the edges of the minor axis that 
is closely aligned with slit position 4 in figure 2 of Sabin et al. (2012). As emphasised in Miszalski et al. (2011b) with NGC~6326, there is
a stark disconnect between the intrinsic morphology of the nebula and the LIS, the latter of which may form separately to the diffuse components.

\section{Conclusions}\label{sec:concl}

We carried out an RV monitoring program of 6 Galactic [WR] CSPNe using a well established cross-correlation method successfully used in massive WR star studies (Foellmi et al. 2003). Four of the CSPNe in our sample are late-types ([WCL]) namely Hen 2-99, Hen 2-113, Hen 3-1333 and PMR 2, and two are early-type ([WCE]), namely the CSPNe of NGC 5189 and NGC 5315. Our main conclusions are as follows:
\begin{enumerate}
 \item No significant variability was detected in Hen 2-113, Hen 3-1333 and PMR 2. Hen 2-99 may be variable with a putative periodicity of 5.3 d, but more observations are needed to further investigate. Similarly, NGC~5315 shows a much greater level of variability but will again require more observations to search for an orbital period. 

\item NGC 5189 showed significant RV variability in stark contrast to the other objects in our sample. A significant period of 4.04 d was found from the Lomb-Scargle periodogram and fitting the phased data gave a peak-to-peak amplitude of $\sim$ 124 km s$^{-1}$. The orbital motion is clearly seen in the trailed diagram of the stellar O~VI 5290 emission line compared against the static H$\beta$ nebular emission line. The mass of the companion $m_2$ has a lower limit of 0.5 M$_\odot$ and may be $\sim$0.9 M$_\odot$ for a reasonable assumption of the nebular inclination of $i=45$ degrees. A more massive WD companion is likely to be present, explaining the lack of an irradiation effect in previous photometric observations of the object. Further spatio-kinematic modelling of NGC~5189 as in Sabin et al. (2012) is strongly encouraged to further constrain the inclination and therefore companion mass.

\item The spectacular nebula of NGC~5189 further fits the trend for post-CE nebulae to be dominated by low-ionisation structures (e.g. NGC~6326, Miszalski et al. 2011) and to possess precessing outflows (e.g. Fleming~1, Boffin et al. 2012) as first outlined by Miszalski et al. (2009b). Several of the PNe with [WR] nuclei discussed in Miszalski et al. (2009b) are therefore excellent candidates for RV monitoring to discover new [WR] binaries. 

\item The discovery of a second close binary system containing a [WR] component strongly suggests mergers are not involved in the formation of most [WC] CSPNe. It remains to be seen whether [WN] CSPNe are binary systems. Indeed, the relatively long orbital period of NGC~5189 could be either a one-off coincidence (e.g. NGC~2346) or alternatively it could indicate that potentially many more [WR] binaries may be found if appropriate RV monitoring surveys are conducted. We speculate that wind interactions between the [WR] component and its companion (e.g. wind Roche-lobe overflow) may be responsible for the longer period. Further RV monitoring of [WR] CSPNe is urged to try and discover more systems to place constraints on the formation of [WR] CSPNe.
\end{enumerate}

\section{Acknowledgements}

We would like to thank Enrico Kotze for producing the trailed diagrams included in this paper. We acknowledge the very helpful discussions and support of Itumeleng Monageng, Rudi Kuhn, Lee Townsend, Paul Crowther, Orsola De Marco and Noam Soker. We thank Prof. R.~H.~M\'endez for his constructive referee report that has helped improve this paper. We are grateful to the SAAO IT team for their wonderful support in terms of computer hardware for the paper write-up. RM thanks the University of Cape Town for financial support to write this paper funded by the paper grant: Postgraduate Publication Incentive (PPI) funding. RM is grateful to the Square Kilometre Array (SKA) and National Research Foundation (NRF) for their financial support of the MSc thesis this work is based on. VM acknowledges financial support from the NRF, South Africa. The observations reported in this paper were obtained with the SAAO 1.9-m telescope and the Southern African Large Telescope (SALT).

\appendix
\clearpage
\newpage
\section{List of close-binaries}
Table \ref{tab:closebinaries} gives a list of published post-CE PNe updating Miszalski et al. (2011c). In the Miszalski et al. (2009a) sample, there are still some objects that require spectroscopic confirmation that the variable star identified is the CSPN. Miszalski (2009) obtained Gemini GMOS spectroscopy for several in the sample. Some objects were removed based partially on the spectra (Miszalski et al. 2011c) and several other objects were confirmed by Miszalski (2009) as true central stars (K~6-34, H~2-29, BMP~1800-3408, PPA~1759-2834, Pe~1-9, PPA~1747-3435, PHR~1757-2824, PHR~1756-3342, Sab~41, M~2-19 and M~3-16). Unpublished spectra of other objects have also been obtained, but these will be discussed elsewhere. We have removed Te~11 which is a cataclysmic variable (Drake et al. 2014) and MPA~1508-6455 which requires further work to prove its long 12.5 d period. A period of 1.26 d is adopted for PN G222.8$-$04.2 based on unpublished SALT RSS radial velocities from our 2013-2-RSA-005 programme.

\renewcommand{\tabcolsep}{0.08cm}
\begin{table*}
   \caption{An updated list of 42 close binary CSPNe whose status is fairly well established.}
   \label{tab:closebinaries}
   \centering
   \begin{tabular}{llrl}
      \hline
      PN G & Name & Period & Discovery reference\\
           &      & (days) &                    \\
      \hline
053.8$-$03.0 & Abell~63 & 0.46 & Bond et al. 1978\\
215.6$+$03.6 & NGC~2346 & 15.99 & Mendez \& Niemela 1981\\
009.6$+$10.5 & Abell~41 & 0.23 & Grauer \& Bond 1983\\
055.4$+$16.0 & Abell~46 & 0.47 & Bond 1985\\
283.9$+$09.7 & DS~1 & 0.36 & Drilling 1985\\
136.3$+$05.5 & HFG~1 & 0.58 & Grauer et al. 1987\\
253.5$+$10.7 & K~1-2 & 0.68 & Bond \& Grauer 1987\\
005.1$-$08.9 & Hf~2-2 & 0.40 & Lutz et al. 2010\\
017.3$-$21.9 & Abell~65 & 1.00 & Bond \& Livio 1990\\
329.0$+$01.9 & Sp~1 & 2.91 & Bond \& Livio 1990\\
355.2$-$03.6 & HaTr~4 & 1.74 & Bond \& Livio 1990\\
144.8$+$65.8 & BE~UMa & 2.29 & Liebert et al. 1995\\
135.9$+$55.9 & SBS~1150+599A & 0.16 & Tovmassian et al. 2004\\
341.6$+$13.7 & NGC~6026 & 0.53 & Hillwig et al. 2010\\
349.3$-$01.1 & NGC~6337 & 0.17 & Hillwig et al. 2010\\
359.1$-$02.3 & M~3-16 & 0.57 & Miszalski et al. 2008\\
357.6$-$03.3 & H~2-29 & 0.24 & Miszalski et al. 2008\\
000.2$-$01.9 & M~2-19 & 0.67 & Miszalski et al. 2008\\
005.0$+$03.0 & Pe~1-9 & 0.14 & Miszalski et al. 2009a\\
355.3$-$03.2 & PPA~1747-3435 & 0.22 & Miszalski et al. 2009a\\
355.7$-$03.0 & H~1-33 & 1.13 & Miszalski et al. 2009a\\
354.5$-$03.9 & Sab~41 & 0.30 & Miszalski et al. 2009a\\
000.6$-$01.3 & Bl~3-15 & 0.27 & Miszalski et al. 2009a\\
359.5$-$01.2 & JaSt~66 & 0.27 & Miszalski et al. 2009a\\
358.7$-$03.0 & K~6-34 & 0.20 & Miszalski et al. 2009a\\
357.0$-$04.4 & PHR~1756-3342 & 0.26 & Miszalski et al. 2009a\\
001.8$-$02.0 & PHR~1757-2724 & 0.80 & Miszalski et al. 2009a\\
001.2$-$02.6 & PHR~1759-2915 & 1.10 & Miszalski et al. 2009a\\
005.0$-$03.1a& MPA~1759-3007 & 0.50 & Miszalski et al. 2009a\\
001.9$-$02.5 & PPA~1759-2834 & 0.31 & Miszalski et al. 2009a\\
357.1$-$05.3 & BMP~1800-3408 & 0.14 & Miszalski et al. 2009a\\
000.9$-$03.3 & PHR~1801-2947 & 0.32 & Miszalski et al. 2009a\\
222.8$-$04.2 & PHR~0654-1045 & 1.26 & Hajduk et al. 2010\\
054.2$-$03.4 & The Necklace & 1.16 & Corradi et al. 2011\\
068.1$+$11.0 & ETHOS~1 & 0.53 & Miszalski et al. 2011a\\
049.4$+$02.4 & Hen~2-428 & 0.18 & Santander-Garc\'ia et al. 2015\\
034.5$-$06.7 & NGC~6778 & 0.15 & Miszalski et al. 2011b\\
338.1$-$08.3 & NGC~6326 & 0.37 & Miszalski et al. 2011b\\
290.5$+$07.9 & Fleming 1 & 1.19 & Boffin et al. 2012\\
259.1$+$00.9 & Hen~2-11 & 0.61 & Jones et al. 2014\\
086.9$-$03.4 & Ou~5 & 0.36 & Corradi et al. 2014\\
307.2$-$03.4 & NGC~5189 & 4.04 & This work\\
      \hline
   \end{tabular}
\end{table*}

\clearpage
\newpage
\section{Log of cross-correlation results}
\renewcommand{\tabcolsep}{0.3cm}
\begin{table*}
   \caption{Log of cross-correlation wavelength ranges, RV shifts, errors and \textsc{xcsao} cross-correlation peak heights $h$.}
\label{tab:appendix}
\begin{tabular}{ccccccc}
\hline
Object&$\lambda$ range & Grating & MJD (mid) & $v_i$  & $\Delta v_i$ & $h$ \\
& (\AA) & &(Days) & (km s${^{-1}}$) & (km s${^{-1}}$) &  \\
\hline
Hen2-113&4477--4705 & 4 &56063.812846 &$-$59.7 & 4.9 & 0.97 \\
                      &&&56063.823819 &$-$60.0 & 4.8 & 0.97 \\
                      &&&56063.836879 &$-$60.1 & 4.7 & 0.97 \\
                      &&&56063.869755 &$-$60.3 & 4.5 & 0.98 \\
                      &&&56064.122231 &$-$56.0 & 4.0 & 0.97 \\
                      &&&56064.801191 &$-$59.0 & 5.0 & 0.97 \\
            &5050--5500&&56063.812846 &$-$61.4 & 7.0 & 0.81 \\
                      &&&56063.823819 &$-$60.0 & 6.0 & 0.82 \\
                      &&&56063.836879 &$-$63.0 & 6.0 & 0.89 \\
                      &&&56063.869755 &$-$60.0 & 6.0 & 0.92 \\
                      &&&56064.122231 &$-$61.1 & 7.2 & 0.90 \\
                      &&&56064.801191 &$-$52.9 & 7.0 & 0.74 \\
          & 5050--5500&6&56069.996481 &$-$50.5 & 5.0 & 0.97 \\
                      &&&56106.823001 &$-$60.1 & 4.9 & 0.98 \\
                      &&&56108.917004 &$-$64.9 & 6.6 & 0.92 \\
                      &&&56108.928125 &$-$58.2 & 5.6 & 0.91 \\
                      &&&56109.813025 &$-$68.5 & 5.7 & 0.89 \\
                      &&&56111.030991 &$-$56.2 & 4.9 & 0.98 \\ 
\hline
Hen3-1333& 4500--4720 &6&56106.927875 &$-$41.8 & 4.4 & 0.98 \\
                      &&&56108.109317 &$-$40.2 & 4.4 & 0.98 \\
                      &&&56109.996351 &$-$48.2 & 4.7 & 0.98 \\
                      &&&56111.057141 &$-$47.3 & 4.3 & 0.99 \\
         &4785--5420&   &56106.927875 &$-$44.8 & 4.1 & 0.98 \\
                      &&&56108.098274 &$-$47.4 & 13.0& 0.58 \\
                      &&&56108.109317 &$-$35.0 & 4.3 & 0.98 \\
                      &&&56109.996351 &$-$47.1 & 4.1 & 0.98 \\
                      &&&56111.057141 &$-$49.1 & 4.6 & 0.97 \\
           &6000--6288 &&56108.899327 &$-$45.0 & 4.0 & 0.98 \\
                      &&&56109.131929 &$-$35.0 & 5.3 & 0.98 \\
                      &&&56109.873401 &$-$45.1 & 3.9 & 0.98 \\
            &6734--7057&&56108.899327 &$-$40.5 & 3.9 & 0.98 \\
                      &&&56109.131929 &$-$41.2 & 4.3 & 0.98 \\
                      &&&56109.873401 &$-$45.1 & 3.9 & 0.98 \\ 
\hline
Hen2-99    &4500--5740&6&56107.745127 &$-$35.6 & 13.7& 0.98 \\
                      &&&56068.820887 &$-$59.0 & 12.7& 0.98 \\
                      &&&56110.976449 &$-$56.0 & 14.3& 0.98 \\
                      &&&56111.821737 &$-$54.5 & 12.6& 0.98 \\
                      &&&56106.744197 &$-$44.5 & 15.0& 0.97 \\
           &6630--7300& &56108.866659 &$-$62.0 & 14.1& 0.96 \\
                      &&&56109.747499 &$-$75.5 & 16.2& 0.96 \\
            &6000--6490&&56108.866659 &$-$58.6 & 15.0& 0.87 \\
\hline

\end{tabular}
\end{table*}

\renewcommand{\tabcolsep}{0.3cm}
\begin{table*}
   \caption{Table \ref{tab:appendix} (continued).}
\label{tab:appendix1}
\begin{tabular}{ccccccc}
\hline
Object  & $\lambda$ range & Grating & MJD (mid) & $v_i$  & $\Delta v_i$ & $h$ \\
& (\AA) &&(Days) & (km s${^{-1}}$) & (km s${^{-1}}$) &  \\
\hline 
NGC 5315&4477--4705&6&56068.848701 & 93.9 & 6.5 & 0.76 \\
                   &&&56106.777051 & 95.8 & 11.5 & 0.83 \\
                   &&&56110.773799 & 65.8 & 6.1 & 0.82 \\
         &5050--5500&&56109.782759 & 79.7 & 6.4 & 0.96 \\
\hline
PMR 2&4477--5400&6&56107.709237 & 2.1 & 6.4& 0.98 \\
                &&&56110.721211 & 4.6 & 12.0& 0.80 \\
                &&&56110.735899 & $-$1.3 & 6.2 & 0.98 \\
                &&&56106.715497 & 0.4 & 6.1 & 0.98 \\
      &6321--6536&&56108.713129 & 9.5 & 6.4 & 0.96 \\
                &&&56109.713509 & 9.7 & 8.5 & 0.94 \\
                &&&56109.843839 & 10.0 & 8.4 & 0.84 \\ 
\hline
NGC 5189&5169--5443&PG 2300&56672.581101 & $-$34.6 & 6.0 & 0.90 \\
                        &&& 56674.596367 & 36.1 & 7.9  & 0.85 \\
                        &&& 56677.556460 & 56.7 & 6.0  & 0.86 \\
                        &&& 56685.583323 & 61.6 & 5.9  & 0.85 \\
                        &&& 56698.625846 & 70.3 & 5.5  & 0.85 \\
                        &&& 56700.550927 & $-$40.4 & 5.6 & 0.87 \\
                        &&& 56701.611668 & 51.6 & 4.4  & 0.90 \\
                        &&& 56704.578300 & $-$37.0 & 6.8 & 0.88 \\
                        &&& 56705.486876 & 49.4 & 3.7  & 0.88 \\
                        &&& 56705.506708 & 44.1 & 6.1  & 0.83 \\
                        &&& 56706.566535 & 57.6 & 5.1  & 0.87 \\
                        &&& 56707.556766 & $-$14.9 & 4.9 & 0.91 \\
                        &&& 56731.589104 & 20.8 & 4.7  & 0.85 \\
                        &&& 56733.414920 & 4.2 & 3.7   & 0.90 \\
\hline
\end{tabular}
\end{table*}

\label{lastpage}

\newpage
\begin{thebibliography}{}
 \bibitem[Acker \& Neiner(2003)]{2003A&A...403..659A} Acker, A., \& Neiner, C.\ 2003, A\&A, 403, 659 
 
 \bibitem[Balick \& Frank(2002)]{2002ARA&A..40..439B} Balick, B., \& Frank, A.\ 2002, \araa, 40, 439 

 \bibitem {} Balick, B. 2007, Asymmetrical Planetary Nebulae IV, http://www.astro.washington. edu/users/balick/PNIC/
 
 \bibitem[Bl{\"o}cker(2001)]{2001Ap&SS.275....1B} Bl{\"o}cker, T.\ 2001, \apss, 275
 \bibitem[Boffin et al.(2012)]{2012Sci...338..773B} Boffin, H.~M.~J., Miszalski, B., Rauch, T., et al.\ 2012, Science, 338, 773 

\bibitem[Bond et al.(1978)]{1978ApJ...223..252B} Bond, H.~E., Liller, W., \& Mannery, E.~J.\ 1978, ApJ, 223, 252 

\bibitem[Bond(1985)]{1985ASSL..113...15B} Bond, H.~E.\ 1985, Cataclysmic Variables and Low-Mass X-ray Binaries, 113, 15 
\bibitem[Bond \& Grauer(1987)]{1987fbs..conf..221B} Bond, H.~E., \& Grauer, A.~D.\ 1987, IAU Colloq.~95: Second Conference on Faint Blue Stars, 221 
\bibitem[Bond \& Livio(1990)]{1990ApJ355..568B} Bond, H.~E., \& Livio, M.\ 1990, ApJ, 355, 568 

 \bibitem[Bond(2000)]{2000ASPC..199..115B} Bond, H.~E.\ 2000, Asymmetrical Planetary Nebulae II: From Origins to Microstructures, 199, 115 
 

 \bibitem[Buckley et al.(2006)]{2006SPIE.6267E..32B} Buckley, D.~A.~H., Swart, G.~P., \& Meiring, J.~G.\ 2006, SPIE, 6267, 32  
 \bibitem[Burgh et al.(2003)]{2003SPIE.4841.1463B} Burgh, E.~B., Nordsieck, K.~H., Kobulnicky, H.~A., et al.\ 2003, SPIE, 4841, 1463 

 \bibitem[Chesneau et al.(2006)]{2006A&A...455.1009C} Chesneau, O., Collioud, A., De Marco, O., et al.\ 2006, A\&A, 455, 1009 



 \bibitem[Ciardullo \& Bond(1996)]{1996AJ....111.2332C} Ciardullo, R., \& Bond, H.~E.\ 1996, \aj, 111, 2332 
 
 \bibitem[Ciardullo et al.(1999)]{1999AJ....118..488C} Ciardullo, R., Bond, H.~E., Sipior, M.~S., et al.\ 1999, AJ, 118, 488 

\bibitem[Cliffe et al.(1995)]{1995ApJ...447L..49C} Cliffe, J.~A., Frank, A., Livio, M., \& Jones, T.~W.\ 1995, ApJL, 447, L49 



 
 \bibitem[Corradi et al.(2011)]{2011MNRAS.410.1349C} Corradi, R.~L.~M., Sabin, L., Miszalski, B., et al.\ 2011, \mnras, 410, 1349 

\bibitem[Corradi et al.(2014)]{2014MNRAS.441.2799C} Corradi, R.~L.~M., Rodr{\'{\i}}guez-Gil, P., Jones, D., et al.\ 2014, MNRAS, 441, 2799 


\bibitem[Cozens et al.(2010)]{2010JAHH...13...59C} Cozens, G., Walsh, A., \& Orchiston, W.\ 2010, Journal of Astronomical History and Heritage, 13, 59 

\bibitem[Crawford et al.(2010)]{2010SPIE.7737E..54C} Crawford, S.~M., Still, M., Schellart, P., et al.\ 2010, Proc. SPIE, 7737E, 54 



 \bibitem[Crowther et al.(1998)]{1998MNRAS.296..367C} Crowther, P.~A., De Marco, O., \& Barlow, M.~J.\ 1998, \mnras, 296, 367 

\bibitem[Crowther et al.(2006)]{2006ApJ...636.1033C} Crowther, P.~A., Morris, P.~W., \& Smith, J.~D.\ 2006, ApJ, 636, 1033 
 
 \bibitem[Crowther(2008)]{2008ASPC..391...83C} Crowther, P.~A.\ 2008, Hydrogen-Deficient Stars, 391, 83 

\bibitem[Davis et al.(2010)]{2010MNRAS.403..179D} Davis, P.~J., Kolb, U., \& Willems, B.\ 2010, MNRAS, 403, 179 

 \bibitem[De Marco \& Soker(2002)]{2002PASP..114..602D} De Marco, O., \& Soker, N.\ 2002, \pasp, 114, 602 
 \bibitem[De Marco et al.(2002)]{2002ApJ...574L..83D} De Marco, O., Barlow, M.~J., \& Cohen, M.\ 2002, \apjl, 574, L83 
\bibitem[De Marco et al.(2004)]{2004ApJ...602L..93D} De Marco, O., Bond, H.~E., Harmer, D., \& Fleming, A.~J.\ 2004, \apjl, 602, L93 
\bibitem[De Marco(2008)]{2008ASPC..391..209D} De Marco, O.\ 2008, Hydrogen-Deficient Stars, 391, 209 
\bibitem[De Marco et al.(2008)]{2008AJ....136..323D} De Marco, O., Hillwig, T.~C., \& Smith, A.~J.\ 2008, AJ, 136, 323 
 \bibitem[De Marco(2009)]{2009PASP..121..316D} De Marco, O.\ 2009, \pasp, 121, 316 
 
\bibitem[Drake et al.(2014)]{2014MNRAS.441.1186D} Drake, A.~J., G{\"a}nsicke, B.~T., Djorgovski, S.~G., et al.\ 2014, MNRAS, 441, 1186 
\bibitem[Drilling(1985)]{1985ApJ...294L.107D} Drilling, J.~S.\ 1985, ApJL, 294, L107 

\bibitem[Evans(1968)]{1968MNSSA..27...37E} Evans, D.~S.\ 1968, Monthly Notes of the Astronomical Society of South Africa, 27, 37 
 \bibitem[Foellmi et al.(2003)]{2003MNRAS.338..360F} Foellmi, C., Moffat, A.~F.~J., \& Guerrero, M.~A.\ 2003, \mnras, 338, 360 
\bibitem[Frew(2008)]{2008PhDT.......109F} Frew, D.~J.\ 2008, Ph.D.~Thesis, Macquarie University
\bibitem[Frew et al.(2014)]{2014MNRAS.440.1345F} Frew, D.~J., Boji{\v c}i{\'c}, I.~S., Parker, Q.~A., et al.\ 2014, MNRAS, 440, 1345 
\bibitem[Garc{\'{\i}}a-Rojas et al.(2012)]{2012A&A...538A..54G} Garc{\'{\i}}a-Rojas, J., Pe{\~n}a, M., Morisset, C., Mesa-Delgado, A., \& Ruiz, M.~T.\ 2012, A\&A, 538, A54 

\bibitem[Gon{\c c}alves et al.(2001)]{2001ApJ...547..302G} Gon{\c c}alves, D.~R., Corradi, R.~L.~M., \& Mampaso, A.\ 2001, \apj, 547, 302 
\bibitem[Gonz{\'a}lez P{\'e}rez et al.(2006)]{2006A&A...454..527G} Gonz{\'a}lez P{\'e}rez, J.~M., Solheim, J.-E., \& Kamben, R.\ 2006, A\&A, 454, 527 
\bibitem[G{\'o}rny(2014)]{2014arXiv1406.1048G} G{\'o}rny, S.~K.\ 2014, A\&A, in press, arXiv:1406.1048 
\bibitem[Grauer \& Bond(1983)]{1983ApJ...271..259G} Grauer, A.~D., \& Bond, H.~E.\ 1983, ApJ, 271, 259 
\bibitem[Grauer et al.(1987)]{1987BAAS...19..643G} Grauer, A.~D., Bond, H.~E., Ciardullo, R., \& Fleming, T.~A.\ 1987, BAAS, 19, 643 
\bibitem[Grosdidier et al.(2000)]{2000A&A...364..597G} Grosdidier, Y., Acker, A., \& Moffat, A.~F.~J.\ 2000, A\&A, 364, 597 
\bibitem[Grosdidier et al.(2001)]{2001ApJ...562..753G} Grosdidier, Y., Moffat, A.~F.~J., Blais-Ouellette, S., Joncas, G., \& Acker, A.\ 2001a, ApJ, 562, 753 
 \bibitem[Guerrero \& Miranda(2012)]{2012A&A...539A..47G} Guerrero, M.~A., \& Miranda, L.~F.\ 2012, \aap, 539, A47 

 \bibitem[Hajduk et al.(2010)]{2010MNRAS.406..626H} Hajduk, M., Zijlstra, A.~A., \& Gesicki, K.\ 2010, \mnras, 406, 626 
\bibitem[Handler(2003)]{2003ASPC..292..183H} Handler, G.\ 2003, Interplay of Periodic, Cyclic and Stochastic Variability in Selected Areas of the H-R Diagram, 292, 183 
 \bibitem[Henize(1967)]{1967ApJS...14..125H} Henize, K.~G.\ 1967, \apjs, 14, 125 
 \bibitem[Herwig(2001)]{2001Ap&SS.275...15H} Herwig, F.\ 2001, \apss, 275, 15 
\bibitem[Hillwig et al.(2010)]{2010AJ....140..319H} Hillwig, T.~C., Bond, H.~E., Af{\c s}ar, M., \& De Marco, O.\ 2010, AJ, 140, 319 
 \bibitem[Hua et al.(1998)]{1998A&AS..133..361H} Hua, C.~T., Dopita, M.~A., \& Martinis, J.\ 1998, \aaps, 133, 361 
\bibitem[Iben \& Livio(1993)]{1993PASP..105.1373I} Iben, I., Jr., \& Livio, M.\ 1993, PASP, 105, 1373 
\bibitem[Ivanova et al.(2013)]{2013A&ARv..21...59I} Ivanova, N., Justham, S., Chen, X., et al.\ 2013, A\&ARv, 21, 59 
 \bibitem[Jones et al.(2014)]{2014A&A...562A..89J} Jones, D., Boffin, H.~M.~J., Miszalski, B., et al.\ 2014, \aap, 562, A89 





\bibitem[Keller et al.(2014)]{2014MNRAS.442.1379K} Keller, G.~R., Bianchi, L., \& Maciel, W.~J.\ 2014, \mnras, 442, 1379 

 \bibitem[Kniazev et al.(2008)]{2008MNRAS.388.1667K} Kniazev, A.~Y., Zijlstra, A.~A., Grebel, E.~K., et al.\ 2008, MNRAS, 388, 1667 

\bibitem[Kobulnicky et al.(2003)]{2003SPIE.4841.1634K} Kobulnicky, H.~A., Nordsieck, K.~H., Burgh, E.~B., et al.\ 2003, SPIE, 4841, 1634 

\bibitem[Koesterke(2001)]{2001Ap&SS.275...41K} Koesterke, L.\ 2001, Ap\&SS, 275, 41 


 
 \bibitem[Kurtz \& Mink(1998)]{1998PASP..110..934K} Kurtz, M.~J., \& Mink, D.~J.\ 1998, \pasp, 110, 934 
 \bibitem[Lagadec et al.(2006)]{2006A&A...448..203L} Lagadec, E., Chesneau, O., Matsuura, M., et al.\ 2006, A\&A, 448, 203 

\bibitem[Lawlor \& MacDonald(2002)]{2002ASPC..279..193L} Lawlor, T.~M., \& MacDonald, J.\ 2002, Exotic Stars as Challenges to Evolution, 279, 193 
\bibitem[Liebert et al.(1995)]{1995ApJ...441..424L} Liebert, J., Tweedy, R.~W., Napiwotzki, R., \& Fulbright, M.~S.\ 1995, ApJ, 441, 424 

 \bibitem[Lomb(1976)]{1976Ap&SS..39..447L} Lomb, N.~R.\ 1976, \apss, 39, 447 

\bibitem[Lopez et al.(1993)]{1993RMxAA..26S.110L} Lopez, J.~A., Roth, M., \& Tapia, M.\ 1993, \rmxaa, 26, 110 
\bibitem[Lutz et al.(2010)]{2010PASP..122..524L} Lutz, J., Fraser, O., McKeever, J., \& Tugaga, D.\ 2010, PASP, 122, 524 



 \bibitem[Markwardt(2012)]{2012ascl.soft08019M} Markwardt, C.\ 2012, Astrophysics Source Code Library, 8019 

\bibitem[Mendez \& Niemela(1981)]{1981ApJ...250..240M} Mendez, R.~H., \& Niemela, V.~S.\ 1981, ApJ, 250, 240 
 \bibitem[Mendez et al.(1988)]{1988A&A...190..113M} Mendez, R.~H., Kudritzki, R.~P., Herrero, A., Husfeld, D., \& Groth, H.~G.\ 1988, A\&A, 190, 113 

 \bibitem[Mendez et al.(1990)]{1990A&A...229..152M} Mendez, R.~H., Herrero, A., \& Manchado, A.\ 1990, \aap, 229, 152 



 \bibitem[Mendez(1991)]{1991IAUS..145..375M} M\'endez, R.~H.\ 1991, Evolution of Stars: the Photospheric Abundance Connection, 145, 375 
\bibitem[Miller Bertolami \& Althaus(2006)]{2006A&A...454..845M} Miller Bertolami, M.~M., \& Althaus, L.~G.\ 2006, A\&A, 454, 845 
 \bibitem[Miller Bertolami et al.(2011)]{2011MNRAS.415.1396M} Miller Bertolami, M.~M., Althaus, L.~G., Olano, C., \& Jim{\'e}nez, N.\ 2011, MNRAS, 415, 1396 

 \bibitem[Miszalski (2009)]{2009M} Miszalski, B.\ 2009, PhD Thesis, Macquarie University and Universit\'e de Strasbourg
 \bibitem[Miszalski et al.(2009)]{2009A&A...496..813M} Miszalski, B., Acker, A., Moffat, A.~F.~J., Parker, Q.~A., \& Udalski, A.\ 2009a, \aap, 496, 813 

 \bibitem[Miszalski et al.(2009)]{2009A&A...505..249M} Miszalski, B., Acker, A., Parker, Q.~A., \& Moffat, A.~F.~J.\ 2009b, \aap, 505, 249 
 
 \bibitem[Miszalski et al.(2011)]{2011MNRAS.413.1264M} Miszalski, B., Corradi, R.~L.~M., Boffin, H.~M.~J., et al.\ 2011a, \mnras, 413, 1264 

\bibitem[Miszalski et al.(2011)]{2011A&A...531A.158M} Miszalski, B., Jones, D., Rodr{\'{\i}}guez-Gil, P., et al.\ 2011b, \aap, 531, A158 

\bibitem [Miszalski et al. (2011)]{} Miszalski, B., R.~L.~M. Corradi, D. Jones, M. Santander-Garc\'ia, P. Rodr\'iguez-Gil, M. M. Rubio-D\'iez\ 2011c, Asymmetric Planetary Nebulae V, arXiv:1009.2890

 \bibitem[Miszalski et al.(2012)]{2012MNRAS.423..934M} Miszalski, B., Crowther, P.~A., De Marco, O., et al.\ 2012, \mnras, 423, 934 

\bibitem[Moe \& De Marco(2006)]{2006ApJ...650..916M} Moe, M., \& De Marco, O.\ 2006, ApJ, 650, 916 


 
\bibitem[Mohamed \& Podsiadlowski(2007)]{2007ASPC..372..397M} Mohamed, S., \& Podsiadlowski, P.\ 2007, 15th European Workshop on White Dwarfs, 372, 397 

\bibitem[Mohamed \& Podsiadlowski(2011)]{2011ASPC..445..355M} Mohamed, S., \& Podsiadlowski, P.\ 2011, Why Galaxies Care about AGB Stars II: Shining Examples and Common Inhabitants, 445, 355 

 \bibitem[Morgan et al.(2001)]{2001MNRAS.322..877M} Morgan, D.~H., Parker, Q.~A., \& Russeil, D.\ 2001, \mnras, 322, 877 

\bibitem[Nagel et al.(2006)]{2006A&A...448L..25N} Nagel, T., Schuh, S., Kusterer, D.-J., et al.\ 2006, A\&A, 448, L25 

\bibitem[Napiwotzki \& Schoenberner(1995)]{1995A&A...301..545N} Napiwotzki, R., \& Schoenberner, D.\ 1995, \aap, 301, 545 
\bibitem {} National Institute of Standard and Technology (NIST), http://www.nist.gov/pml/data/asd.cfm, [Online; accessed in 2013]

\bibitem[Nebot G{\'o}mez-Mor{\'a}n et al.(2011)]{2011A&A...536A..43N} Nebot G{\'o}mez-Mor{\'a}n, A., G{\"a}nsicke, B.~T., Schreiber, M.~R., et al.\ 2011, A\&A, 536, A43 
\bibitem[O'Donoghue et al.(2006)]{2006MNRAS.372..151O} O'Donoghue, D., Buckley, D.~A.~H., Balona, L.~A., et al.\ 2006, MNRAS, 372, 151 

\bibitem[Osuna et al.(2005)]{2005ASPC..347..198O} Osuna, P., Barbarisi, I., Salgado, J., \& Arviset, C.\ 2005, Astronomical Data Analysis Software and Systems XIV, 347, 198 

\bibitem[Phillips(1984)]{1984A&A...137...92P} Phillips, J.~P.\ 1984, \aap, 137, 92 

\bibitem[Phillips \& Reay(1983)]{1983A&A...117...33P} Phillips, J.~P., \& Reay, N.~K.\ 1983, \aap, 117, 33 

\bibitem[Press \& Rybicki(1989)]{1989ApJ...338..277P} Press, W.~H., \& Rybicki, G.~B.\ 1989, \apj, 338, 277 

\bibitem[Raga et al.(2009)]{2009ApJ...707L...6R} Raga, A.~C., Esquivel, A., Vel{\'a}zquez, P.~F., et al.\ 2009, ApJL, 707, L6 

 
\bibitem[Rauch et al.(1994)]{1994A&A...286..543R} Rauch, T., K\"oppen, J., \& Werner, K.\ 1994, A\&A, 286, 543 
\bibitem[Rauch et al.(1996)]{1996A&A...310..613R} Rauch, T., K\"oppen, J., \& Werner, K.\ 1996, A\&A, 310, 613 
\bibitem[Rauch et al.(1998)]{1998A&A...338..651R} Rauch, T., Dreizler, S., \& Wolff, B.\ 1998, A\&A, 338, 651 
\bibitem[Rauch et al.(2008)]{2008ASPC..391..135R} Rauch, T., Reiff, E., Werner, K., \& Kruk, J.~W.\ 2008, Hydrogen-Deficient Stars, 391, 135 
\bibitem[Rebassa-Mansergas et al.(2008)]{2008MNRAS.390.1635R} Rebassa-Mansergas, A., G{\"a}nsicke, B.~T., Schreiber, M.~R., et al.\ 2008, MNRAS, 390, 1635 
   

\bibitem[Reindl et al.(2013)]{2013ASPC..469..143R} Reindl, N., Rauch, T., Werner, K., \& Kruk, J.~W.\ 2013, 18th European White Dwarf Workshop., 469, 143 
\bibitem[Reindl et al.(2014)]{2014A&A...566A.116R} Reindl, N., Rauch, T., Werner, K., Kruk, J.~W., \& Todt, H.\ 2014, A\&A, 566, A116 
\bibitem[Sabin et al.(2012)]{2012RMxAA..48..165S} Sabin, L., V{\'a}zquez, R., L{\'o}pez, J.~A., Garc{\'{\i}}a-D{\'{\i}}az, M.~T., \& Ramos-Larios, G.\ 2012, \rmxaa, 48, 165 

 \bibitem[Sana et al.(2013)]{2013MNRAS.432L..26S} Sana, H., van Boeckel, T., Tramper, F., et al.\ 2013, \mnras, 432, L26 
 \bibitem[SG2015]{2015Natur} Santander-Garc\'ia, M., Rodr{\'{\i}}guez-Gil, P., Corradi, R.L.M., Jones, D., Miszalski, B., Boffin, H.M.J., Rubio-D\'iez, M.M., \& Kotze, M.M., 2015, Nature, in press

 \bibitem[Scargle(1982)]{1982ApJ...263..835S} Scargle, J.~D.\ 1982, \apj, 263, 835 

 \bibitem[Schoenberner(1979)]{1979A&A....79..108S} Schoenberner, D.\ 1979, \aap, 79, 108
 
 \bibitem[Schnurr(2008)]{2008PhDT........34S} Schnurr, O.\ 2008, Ph.D.~Thesis,  

 \bibitem[Schnurr et al.(2009)]{2009MNRAS.395..823S} Schnurr, O., Moffat, A.~F.~J., Villar-Sbaffi, A., St-Louis, N., \& Morrell, N.~I.\ 2009, \mnras, 395, 823 

\bibitem[Schuh et al.(2009)]{2009JPhCS.172a2065S} Schuh, S., Beeck, B., \& Nagel, T.\ 2009, Journal of Physics Conference Series, 172, 012065 

 
 \bibitem[Smith et al.(1996)]{1996MNRAS.281..163S} Smith, L.~F., Shara, M.~M., \& Moffat, A.~F.~J.\ 1996, \mnras, 281, 163



    

\bibitem[Stanghellini et al.(2008)]{2008ApJ...689..194S} Stanghellini, L., Shaw, R.~A., \& Villaver, E.\ 2008, ApJ, 689, 194 


    

\bibitem[Tanner(1948)]{1948JRASC..42..177T} Tanner, R.~W.\ 1948, JRASC, 42, 177 


 
 
 \bibitem[Todt et al.(2010)]{2010A&A...515A..83T} Todt, H., Pe{\~n}a, M., Hamann, W.-R., \& Gr{\"a}fener, G.\ 2010, \aap, 515, A83 
 
 \bibitem[Todt et al.(2013)]{2013MNRAS.430.2302T} Todt, H., Kniazev, A.~Y., Gvaramadze, V.~V., et al.\ 2013, \mnras, 430, 2302 


 \bibitem[Tonry \& Davis(1979)]{1979AJ.....84.1511T} Tonry, J., \& Davis, M.\ 1979, \aj, 84, 1511 
\bibitem[Tovmassian et al.(2004)]{2004ApJ...616..485T} Tovmassian, G.~H., Napiwotzki, R., Richer, M.~G., et al.\ 2004, ApJ, 616, 485 

\bibitem[Trumpler \& Weaver(1953)]{1953stas.book.....T} Trumpler, R.~J., \& Weaver, H.~F.\ 1953, Dover Books on Astronomy and Space Topics, New York: Dover Publications, |c1953,  



 
 \bibitem[van der Hucht(2001)]{2001NewAR..45..135V} van der Hucht, K.~A.\ 2001, \nar, 45, 135 

\bibitem[van Dokkum(2001)]{2001PASP..113.1420V} van Dokkum, P.~G.\ 2001, \pasp, 113, 1420 
\bibitem[van Winckel et al.(2009)]{2009A&A...505.1221V} van Winckel, H., Lloyd Evans, T., Briquet, M., et al.\ 2009, A\&A, 505, 1221 

\bibitem[Werner \& Herwig(2006)]{2006PASP..118..183W} Werner, K., \& Herwig, F.\ 2006, \pasp, 118, 183 

\bibitem[Werner et al.(2008)]{2008ASPC..391..109W} Werner, K., Rauch, T., Reiff, E., \& Kruk, J.~W.\ 2008, Hydrogen-Deficient Stars, 391, 109 

\bibitem[Werner(2012)]{2012IAUS..283..196W} Werner, K.\ 2012, IAU Symposium, 283, 196 

\bibitem[Zhang \& Jeffery(2012)]{2012MNRAS.419..452Z} Zhang, X., \& Jeffery, C.~S.\ 2012, MNRAS, 419, 452 

\bibitem[Zijlstra et al.(2006)]{2006MNRAS.369..875Z} Zijlstra, A.~A., Gesicki, K., Walsh, J.~R., et al.\ 2006, MNRAS, 369, 875 

\end{thebibliography}
\end{document}